\documentclass[11pt]{article}
\usepackage{amsmath,amssymb,color,graphics,epsfig}

\usepackage{amsfonts,epsf,graphicx}

\makeatletter
\@addtoreset{equation}{section}
\makeatother

\usepackage{amsfonts}

\newcommand{\hoch}[1]{$\, ^{#1}$}

\newcommand{\be}{\begin{equation}}
\newcommand{\ee}{\end{equation}}
\newcommand{\bea}{\setlength\arraycolsep{2pt} \begin{eqnarray}}
\newcommand{\eea}{\end{eqnarray}}
\newcommand{\nn}{\nonumber}

\def\ft#1#2{{\textstyle{\frac{\scriptstyle #1}{\scriptstyle #2} } }}
\def\fft#1#2{{\frac{#1}{#2}}}

\def\0{{\sst{(0)}}}
\def\1{{\sst{(1)}}}
\def\2{{\sst{(2)}}}
\def\3{{\sst{(3)}}}
\def\4{{\sst{(4)}}}
\def\5{{\sst{(5)}}}
\def\6{{\sst{(6)}}}
\def\7{{\sst{(7)}}}
\def\8{{\sst{(8)}}}
\def\sst#1{{\scriptscriptstyle #1}}

\def\ep{{\epsilon}}

\begin{document}

\begin{flushright}
\end{flushright}

\vspace{15pt}
\begin{center}
{\large {\bf Strings on AdS Wormholes and Nonsingular Black Holes}}

\vspace{10pt}
H. L\"u\hoch{1}, Justin F. V\'azquez-Poritz\hoch{2,3} and Zhibai Zhang\hoch{2,3}

\vspace{10pt}

\hoch{1}{\it Department of Physics, Beijing Normal University,
Beijing 100875, China}

\vspace{10pt}

\hoch{2}{\it Physics Department\\ New York City College of Technology, The City University of New York\\ 300 Jay Street, Brooklyn NY 11201, USA}

\vspace{10pt}

\hoch{3}{\it The Graduate School and University Center, The City University of New York\\ 365 Fifth Avenue, New York NY 10016, USA}\\

\vspace{30pt}

\underline{ABSTRACT}
\end{center}

Certain AdS black holes in the STU model can be conformally scaled to wormhole and black hole backgrounds which have two asymptotically AdS regions and are completely free of curvature singularities. While there is a delta-function source for the dilaton, classical string probes are not sensitive to this singularity. According to the AdS/CFT correspondence, the dual field theory lives on the union of the disjoint boundaries. For the wormhole background, causal contact exists between the two boundaries and the structure of certain correlation functions is indicative of an interacting phase for which there is a coupling between the degrees of freedom living at each boundary. The nonsingular black hole describes an entangled state in two non-interacting identical conformal field theories. By studying the behavior of open strings on these backgrounds, we extract a number of features of the ``quarks'' and ``anti-quarks'' that live in the field theories. In the interacting phase, we find that there is a maximum speed with which the quarks can move without losing energy, beyond which energy is transferred from a quark in one field theory to a quark in the other. We also compute the rate at which moving quarks within entangled states lose energy to the two surrounding plasmas. While a quark-antiquark pair within a single field theory exhibits Coulomb interaction for small separation, a quark in one field theory exhibits spring-like confinement with an anti-quark in the other field theory. For the entangled states, we study how the quark-antiquark screening length depends on temperature and chemical potential.

\thispagestyle{empty}

\pagebreak



\newpage

\section{Introduction}

The AdS/CFT correspondence \cite{ads-cft1,ads-cft2,ads-cft3} enables us to extract information on certain strongly-coupled gauge theories by studying string theory on a supergravity background. While the original correspondence was between type IIB string theory on AdS$_5\times S^5$ and four-dimensional maximally supersymmetric ${\cal N}=4$ $SU(N)$ Yang-Mills theory, a large number of generalizations have been made (see e.g.~\cite{agmoo}). The field theory interpretation of various types of geometrical backgrounds have been investigated, including those with two disconnected asymptotically AdS regions such as the eternal black hole \cite{eternal} and the Maldacena-Maoz wormhole \cite{maoz}. The correspondence may be a powerful tool for addressing long-standing questions in quantum gravity. An indication of this is that causal properties of the bulk spacetime, such as the formation of an event horizon and the presence of a black hole singularity, are manifest in the dual field theory in terms of the singular structure of correlation functions \cite{sensitive1,sensitive2,sensitive3}.

The three-charge AdS black hole solution in the STU model \cite{Behrndt1,Behrndt2} has found widespread application in the study of the AdS/CFT correspondence. In particular, this gravity background is conjectured to be dual to non-zero temperature ${\cal N}=4$ $SU(N)$ supersymmetric Yang-Mills theory with chemical potentials for the $U(1)$ $R$-charges \cite{Cvetic-Gubser}. For static black holes in the STU model, the BPS limit leads to naked singularities \cite{Behrndt1}. These BPS naked singularities, which are also called `superstars,' were shown to correspond to a distribution of giant gravitons in AdS$_5\times S^5$ \cite{superstars}. If one of the charges vanishes, then there is a conformal scaling for which the geometry can be extended to have two asymptotic AdS regions and be completely free of curvature singularities. This is the so-called dual frame, which is the frame in which the dual field strength couples to the dilaton in the same manner as the metric \cite{dual-frame}. If we were in ten dimensions, then the dual frame may be considered as a ``holographic frame" \cite{skenderis}. However, since we are in five dimensions, we can at most consider this to be a background for an effective five-dimensional string theory.

Depending on the values of the parameters, there might or might not be horizons present, so that the background is either a wormhole or a nonsingular black hole. We use the term ``nonsingular black hole" to indicate the absence of a geometrical singularity in a solution with event horizon although, as we will see in a moment, this background is supported by a delta function source for the dilaton. There have been a number of approaches to the construction of nonsingular black holes; see \cite{bardeen} for an early example and \cite{anabalon} for a recent one. We will focus on the case in which the two remaining charges are equal. Then the dilaton becomes non-dynamical in the dual frame. In analogy with $f(R)$ gravity (see \cite{review1,review2,review3} for reviews of $f(R)$ theories) being equivalent to a special class of Brans-Dicke theory in which the scalar field has no kinetic term, this can also be referred to as the ``$f(R)$ frame'' \cite{susyfr}. Although the background has no curvature singularities, it is supported by a delta function source for the dilaton. Nevertheless, matter that is not coupled to the dilaton through a derivative will not be sensitive to this singularity. In particular, this is the case for a classical string probe.

We will make the working assumptions that a five-dimensional gauged supergravity solution in the $f(R)$ frame can be used as a background for an effective string theory in five dimensions and that the AdS/CFT correspondence can be applied in this context. Then the dual field theory lives on the union of the disjoint boundaries \cite{Witten-Yau}. In the case of the wormhole background, causal contact exists between the two boundaries. We compute correlation functions between fields living on different boundaries. Their structure is indicative of interactions between the degrees of freedom living at each boundary. On the other hand, the nonsingular black hole describes an entangled state in two copies of conformal field theories. A string that connects $N$ branes to a single brane in an asymptotic region corresponds to an infinitely massive W-boson, which we will refer to as a ``quark'' of one of the field theories-- which field theory is determined by which asymptotic region. We will consider the dynamics of open strings on this background, which will enable us to extract a number of features of the field theories.

This paper is organized as follows. In section 2, we discuss how the STU model with two equal field strengths can be converted to the $f(R)$ frame by performing a conformal scaling and then integrating out the scalar. Taking a two-equal-charge AdS black hole to the $f(R)$ frame renders the geometry completely regular with two asymptotic AdS regions. In section 3, we discuss the dual field theories and compute correlation functions. For the wormhole background, the structure of the correlation functions indicates that the degrees of freedom associated with different boundaries are coupled. In section 4, we study the behavior of steadily-moving strings on this background. In the interacting phase, we find that there is a maximum speed with which the quarks can move without losing energy, beyond which energy is transferred from a quark in one field theory to a quark in the other. We also compute the rate at which moving quarks within entangled states lose energy to the two surrounding plasmas.

In section 5, we analyze the interaction potential of quark-antiquark pairs in these field theories. While a quark-antiquark pair within a single field theory exhibits Coulomb interaction for small separation, a quark in one field theory exhibits spring-like confinement with an anti-quark in the other field theory. For the entangled states, we study how the quark-antiquark screening length depends on temperature and chemical potential. In the interacting phase of the two field theories, a quadruplet made up of one quark-antiquark pair in each field theory can undergo transitions involving the pairing for screening. Namely, on one side of the transition the quarks interact with the antiquarks of the same field theory and are screened from the antiquarks of the other field theory, while it is the reverse on the other side of the transition. We give the conclusions in section 6. In an appendix, the generalization of these backgrounds to arbitrary dimensionality is presented, along with a discussion of the global structure and the thermodynamical quantities.

\section{Obtaining the backgrounds}

\subsection{The $f(R)$ frame}

$f(R)$ gravity is equivalent to a special class of the Brans-Dicke theory in which the scalar field has no kinetic term. Conversely, any gravity theory coupled to a scalar field can be conformally scaled into the ``$f(R)$ frame'' in which the scalar has no kinetic term and may become auxiliary. Integrating out this auxiliary field then gives rise to an $f(R)$ theory. In analogy with this, one can conformally scale supergravity theories to the $f(R)$ frame, such that the dilaton becomes an auxiliary field and can be integrated out \cite{susyfr}.

We start off by considering the STU model, which corresponds to five-dimensional ${\cal N}=2$ gauged supergravity coupled to two vector multiplets. If we set two field strengths equal, then we have the Lagrangian
\begin{eqnarray}
e^{-1}{\cal L}_5 &=& R - \ft12(\partial \phi)^2 +4g^2 \Big(2
e^{-\fft{1}{\sqrt6} \phi} + e^{\fft2{\sqrt6} \phi}\Big) -
\ft14e^{-\fft{2}{\sqrt6}\phi} F_\2^2 - \ft14 e^{\fft4{\sqrt6}\phi}
{\cal F}_\2^2\cr 
&& +\ft18 e^{-1}\epsilon^{\mu\nu\rho\sigma\lambda} F_{\mu\nu}
F_{\rho\sigma} {\cal A}_\lambda\,,
\end{eqnarray}
where $F_\2=dA_\1$ and ${\cal F}_\2=d{\cal A}_\1$. Here we have adopted the notation of \cite{tenauthors} where the five-sphere reduction ansatz of the STU model from type IIB supergravity were given. Upon making the conformal transformation and field redefinition for the dilaton
\be
G_{\mu\nu}\rightarrow e^{\fft{\phi}{\sqrt{6}}} G_{\mu\nu}\,,\qquad \varphi=e^{\sqrt{\fft{3}{8}} \phi}\,,
\ee
the Lagrangian becomes \cite{susyfr}
\begin{equation}
e^{-1} {\cal L}_5 = \varphi (R + 8g^2) + \varphi^3 (4g^2 - \ft14 {\cal
F}_\2^2) - \ft14 \varphi^{-1} F_\2^2 - \ft14 e^{-1}
\epsilon^{\mu\nu\rho\sigma\lambda} F_{\mu\nu} F_{\rho\sigma} {\cal
A}_\lambda\,.\label{d5suplag1}
\end{equation}
In this so-called $f(R)$ frame, the dilaton $\varphi$ is non-dynamical, giving a purely
algebraic equation for $\varphi$:
\begin{equation}
R + 8g^2 + 3\varphi^2 (4g^2 - \ft14 {\cal F}_\2^2) +
\fft{1}{4\varphi^2} F_\2^2=0\,.
\end{equation}

An alternative route for obtaining the Lagrangian given by (\ref{d5suplag1}) is to start with ${\cal N}=(1,0)$ supergravity in six dimensions, which is the low-energy effective action of the self-dual string. Performing an $S^1$ reduction in the string frame, or the six-dimensional frame, namely
\begin{equation}
ds_6^2=ds_5^2 + \varphi^2 (dz + {{\cal A}_{(1)}})^2\,,
\end{equation}
and gauging the resulting five-dimensional theory yields the theory described by (\ref{d5suplag1}). Before the gauging, it can be shown that the $f(R)$ frame in five dimensions corresponds to the dual frame, or equivalently an effective string frame, by dualizing the 2-form field strength $F_\2$ to be the three-form, namely
\begin{equation}
H_\3=\varphi^{-1} {*F_\2}\,.\label{dualize}
\end{equation}
Then the Lagrangian given by (\ref{d5suplag1}) takes the form
\begin{equation}
e^{-1} {\cal L}_5=\varphi\left( R - \ft1{12} H_\3^2\right) + \cdots\,,
\end{equation}
which is characteristic for a theory in the dual frame \cite{dual-frame}. In fact, the $H_\3$ is the direct descendent of the self-dual 3-form in the six-dimensional self-dual string.

\subsection{Wormholes and non-singular black holes}

It should be emphasized that the theory in the $f(R)$ frame is not necessarily equivalent to the original supergravity theory, since the conformal scaling can be singular in the solution space. Therefore, the theory in the $f(R)$ frame can have solutions with global properties that differ from those of the corresponding local solutions of the supergravity theory. In particular, previously singular geometries can become regular in the $f(R)$ frame.  The dilaton remains singular but it is associated with a delta-function source in the $f(R)$ frame.  The theory distinguishes itself also by how it couples with the matter fields.  It is clear that, in the dual frame, a classical string probe will not be sensitive to the delta-function source of the dilaton.

As our starting point, we will consider the static three-charge AdS black hole solution in the STU model \cite{Behrndt1,Behrndt2}. The BPS limit leads to naked singularities known as `superstars,' which correspond to a distribution of giant gravitons in AdS$_5\times S^5$ \cite{superstars}. However, for vanishing ${\cal F}_\2$ there is a conformal transformation which renders the geometry completely regular. For the case of two equal charges, such a conformal transformation takes the black hole solution to the $f(R)$ frame. The resulting solution is given by
\begin{eqnarray}\label{solution2}
ds_5^2 &=& - H^{-1} h dt^2 + H \Big(\fft{dr^2}{h} + r^2
d\Omega_{3,\ep}^2\Big)\,,\nn\\
F_\2 &=& \sqrt{2(\ep + \mu q^{-1})}\, dt\wedge dH^{-1}\,,\qquad \varphi =\fft{|r|}{\sqrt{r^2 +
 q}}\,,
\end{eqnarray}
where
\be
h = \ep -\fft{\mu}{r^2}+ g^2 r^2 H^2\,,\qquad H=1 + \fft{q}{r^2}\,.
\ee
Here $d\Omega^2_{3,\ep}$ is the metric for the space $\Omega^{3,\ep}$, which is a unit sphere $S^3$, flat plane ${\mathbb R}^3$ or hyperbolic plane $H^3$ for $\ep=1,0$ and $-1$, respectively.

The geometry has two asymptotic AdS regions at $r \rightarrow -\infty$ and $r \rightarrow +\infty$ and is completely free of curvature singularities for nonzero $q$. In particular, the scalar curvature invariants go as $1/(r^2+q)^n$ for some $n$ and are everywhere finite. In the region around $r=0$, the geometry approaches a direct product of two-dimensional Minkowski spacetime and $\Omega^{3,\ep}$, or five-dimensional Minkowski spacetime for $\ep=0$. The absolute value on $r$ in (\ref{solution2}) is added in by hand so that $\varphi$ is non-negative in order to avoid ghost fields.  Thus, the solution has a delta-function source at $r=0$. However, any matter that is not coupled to $\varphi$ through a derivative, such as a classical string probe, will not be sensitive to this $\delta$-function singularity.

For $\mu<g^2q^2$, the solution is a wormhole. There are no horizons and the two asymptotic regions are causally connected. In particular, as we will discuss in more detail for the case of $\ep=0$, a radial null geodesic goes from one asymptotic region to the other in a finite time interval. The topological censorship theorem that the disconnected boundaries must be separated by horizons assumes that the Einstein equations hold \cite{galloway}, and therefore does not necessarily carry over to $f(R)$ type theories. Note that the parameter $\mu$ has no lower bound.

For $\mu>g^2q^2$, there are event horizons located at $r=\pm r_h$, where
\be
r_h=\fft{1}{g} \sqrt{\sqrt{g^2\mu+\ep g^2q+\fft{\ep^2}{4}}-\left( g^2q+\fft{\ep}{2}\right)}\,.
\ee
As one might expect, $t$ becomes a spacelike coordinate and $r$ becomes a timelike coordinate in the region between the horizons. $\mu=g^2q^2$ corresponds to the extremal limit for which the near-horizon geometry is AdS$_2\times\Omega^{3,\ep}$. 

For the rest of the paper, we will focus on the case of $\ep=0$. After performing the rescaling
\be\label{rescaling}
r=\ell^2 \rho,\qquad d\Omega_{3,\ep}^2\rightarrow \ell^{-2} d\Omega_{3,\ep}^2,\qquad q\rightarrow \ell^4 q,\qquad \mu\rightarrow \ell^6 \mu\,,
\ee
where $\ell=g^{-1}$, the metric in (\ref{solution2}) for $\ep=0$ can be written as
\be\label{metric}
ds_5^2 = \ell^2 \left( -fdt^2+f^{-1} d\rho^2+(\rho^2+q) (dx^2+dy^2+dz^2)\right)\,,
\ee
where
\be
f=\fft{(\rho^2+q)^2-\mu}{\rho^2+q}\,.
\ee

For the nonsingular black hole, the temperature and entropy density are given by
\be\label{temperature}
T=\fft{1}{\pi} \sqrt{\sqrt{\mu}-q}\,.
\ee
and
\be\label{entropy}
s=\fft{\ell^3 \sqrt{\mu}\sqrt{\sqrt{\mu}-q}}{4G_N}\,,
\ee
respectively. Unlike for dilatonless charged black holes, in this case the entropy vanishes in the extremal limit $\mu=q^2$, where we are using the relation between the parameters after the rescaling (\ref{rescaling}) has been performed. 

The mass and two equal-valued charges are
\be
M=\fft{3\ell^3 \mu}{16\pi G_N}\,,\qquad Q=\fft{\ell^2 \sqrt{2\mu q}}{8\pi G_N}\,,
\ee
and the first law $dM=TdS + \Phi dQ$ is satisfied. The physical charges of the black hole correspond to $R$-charges in the dual field theories. The $R$-charge chemical potentials of the field theory are equated with the electric potentials at the horizons, both of which are given by
\be
\Phi=\ell\sqrt{2q}\,.
\ee

\section{AdS/CFT and correlation functions}

According to the AdS/CFT correspondence, the partition function for string theory on AdS (which reduces to the on-shell action in the semiclassical limit) can be interpreted as a functional of boundary data and equated to the generating function of correlation functions of the dual field theory \cite{ads-cft2,ads-cft3}. For gravitational backgrounds with two asymptotic AdS regions, the usual prescription for holography would imply that the dual field theory lives on the union of the disjoint boundaries. Given that the boundaries are disconnected, one might expect that the field theory is simply the product of the theories on each one of the boundaries, and it is not apparent how the theories on the different manifolds could be coupled \cite{Witten-Yau}. 
However, for the wormhole background, the causal connection between the boundaries leads to an interaction term between the two sets of degrees of freedom living at each boundary \cite{orbifold}. 

The two boundaries are associated with the existence of two sets of dual operators $\mathcal{O}_{1,2}$ corresponding to the two independent boundary conditions $\phi_0^{1,2}$ that must be imposed on a field $\phi$ when solving the wave equation. We will apply the well-known prescription for computing correlation functions of a dual conformal field theory in terms of the bulk data, which describe observables in the asymptotic spatial boundary regions. Non-vanishing results for correlators between operators located at different boundaries indicate that there is indeed a coupling between the fields associated with each boundary \cite{arias}. Therefore, the wormhole background corresponds to a dual field theory with two copies of fundamental fields that interact.

For $\mu<g^2q^2$, there are no horizons and the two asymptotic regions are causally connected. It can be shown radial null geodesic goes from one asymptotic region to the other in a finite time interval, which is given by
\be
\Delta t=\fft{\pi}{2}\left( \fft{1}{\sqrt{q+\sqrt{\mu}}}+\fft{1}{\sqrt{q-\sqrt{\mu}}}\right)\,,
\ee
as measured by an observer in an asymptotic region. 

Suppose that we have a scalar field of mass $m$ in the bulk, which is dual to an operator in the field theory whose 2-point correlation function we wish to compute. We will use the AdS/CFT prescription for determining 2-point correlators from a computation of the bulk propagator. For large mass bulk fields, the correlation functions can be evaluated in the semiclassical geodesic approximation. The bulk propagator is obtained by summing over paths between two points in the bulk, where each path is weighted by $e^{-m{\cal L}}$ and ${\cal L}$ is the suitably regulated proper length of the path. In the large mass limit, the bulk propagator is dominated by the shortest geodesic connecting the two points. 

As an example, consider the wormhole background for which the parameter $\mu=0$. Then we find that the 2-point correlation function between operators on the same boundary goes as
\be\label{correlator1}
\left< {\cal O}_i {\cal O}_i\right> (t,s) \sim \fft{1}{\sin^{2m} \left( \fft{\sqrt{q}}{2} \sqrt{t^2-s^2}\right)}\,,
\ee
where $t$ is the difference in time and $s$ is the distance between the two points along the boundary directions. The index $i=1$ or $2$ denotes the boundary on which the operators live. Note that the correlation function has the expected conformal behavior $(t^2-s^2)^{-m}$ when the operators approach each other. The correlator (\ref{correlator1}) has singular points which reflect the existence of null geodesics going back and forth between the two points. This can be seen most clearly for $s=0$, in which case the singularities in the correlator occur at times $2n\pi$, which is the time it takes for a radial null geodesic to make $n$ roundtrips between the two boundaries, where $n$ is an integer.

For operators on different boundaries, the 2-point correlation function goes as
\be\label{correlator2}
\left< {\cal O}_1 {\cal O}_2\right> (t,s) \sim \fft{1}{\cos^{2m} \left( \fft{\sqrt{q}}{2} \sqrt{t^2-s^2}\right)}\,,
\ee
As before, the singularities of the correlator are due to null geodesics. In particular, for $s=0$ the singularities occur at times $(2n+1)\pi$, which is the time it takes for a radial null geodesic to go back and forth any number of times, beginning at one boundary and ending at the other. It has been argued that the existence of the singularities for the correlator (\ref{correlator2}) is associated with an interaction between the degrees of freedom at the different boundaries \cite{orbifold}. This interpretation assumes that the field theories associated with each boundary live on the same spacetime.

One can perturbatively expand the correlation functions in terms of $\mu$ such that $\mu\ll q^2$. For instance, for operators on different boundaries but with $s=0$ we find
\bea
\left< {\cal O}_1 {\cal O}_2\right> (t) &\sim& \fft{1}{\cos^{2m} \left( \fft{\sqrt{q}}{2} t\right)}\nn\\
&\times&  \left[ 1-\fft{\mu}{8q^2} \csc^2 \left( \fft{\sqrt{q}}{2} t\right) \bigg( 5+\cos\left( 2\sqrt{q} t\right) -6\sqrt{q} t \cot \left( \sqrt{q} t \right)\bigg)+\cdots \right]^m.
\eea
This perturbative expansion breaks down at times close to $\ft{n\pi}{\sqrt{q}}$, where $n$ is any integer.
Since the boundary metric exhibits Lorentz symmetry for $\mu=0$, up to the leading order correction one can introduce a spatial separation $s$ along the boundary directions by taking $t\rightarrow \sqrt{t^2-s^2}$.

We will now consider the nonsingular black hole background, for which $\mu>g^2q^2$ and the asymptotic regions are separated by two event horizons. Like the eternal AdS Schwarzschild black hole \cite{eternal}, we propose that this describes an entangled state in two identical decoupled conformal field theories given by
\be
\left| \Psi\right> =\sum_n e^{-\fft{\beta}{2} E_n} \left| E_n\right>_1\times \left| E_n\right>_2\,,
\ee
where $\left| E_n\right>_{1,2}$ denotes an energy eigenstate in the two conformal field theories and $\beta=T^{-1}$, where $T$ is given by (\ref{temperature}). Correlation functions involving only operators in one of the conformal field theories in this entangled state lead to thermal correlation functions
\be
\left<\Psi\left| \mathcal{O}_1\right|\Psi\right> =\sum_n e^{-\beta E_n}\ _1\left< E_n\left| \mathcal{O}_1\right| E_n\right>_1 =\text{Tr} \left( \rho_{\beta} \mathcal{O}_1\right)\,,
\ee
where $\rho_{\beta}$ is the thermal density matrix. The entropy of entanglement is given by the entropy of the black hole (\ref{entropy}).

The coordinate time that an infalling radial null geodesic takes to go from one of the asymptotic regions to $r=0$ is
\be
\Delta t=\fft{\pi}{4 \sqrt{q+\sqrt{\mu}}}-\fft{i}{4T}\,,
\ee
where the form of the imaginary time piece in terms of the temperature $T$ is a generic feature for black holes. Unlike the eternal AdS Schwarzschild black hole, this is a nonsingular black hole in that there is no curvature singularity. The presence of a horizon as well as the absence of a curvature singularity can be detected in the field theories by examining the structure of the singularities of boundary correlation functions \cite{sensitive1,sensitive2,sensitive3}. Since this stems from the behavior of geodesics, one can roughly trace the results back to the characteristics of the effective potential for a particle on the geometry. For instance, for the eternal AdS Schwarzschild black hole, a divergence in the effective potential associated with the curvature singularity leads to a singularity in a correlator when it is analytically continued past a branch cut to a secondary sheet. For the nonsingular black hole, the effective potential does not diverge and so this branch cut does not arise. On the other hand, for both black hole backgrounds the effective potential has maxima which correspond to the existence of unstable circular orbits. This ultimately leads to singularities in correlators that signal the presence of horizons.

As with the eternal black hole \cite{eternal}, the nonsingular black hole background can be interpreted as two maximally entangled black holes. It has been claimed that any pair of entangled black holes are connected by an Einstein-Rosen bridge, which constitutes a manifestation of the entanglement \cite{EPR=ER}. The nonsingular black hole background does provide an example that is in line with this conjectured relation. However, this is for a special case in which the spacetime behind the horizons of the black holes does not need to be filled in with an Einstein-Rosen bridge, since there is already a smooth passage from the interior of one black hole to that of the other. In particular, the geometry around the central region is that of five-dimensional Minkowski spacetime. If a signal is sent into one black hole, then it will pass through to the interior region of the other black hole and behave as though it is in a quantum well. In particular, such a signal could be used to create a firewall in the interior region of the other black hole. A more energetic signal will lead to a firewall that is all the more closer to the horizon.

\section{Steadily-moving strings}

\subsection{General formulae}

The dynamics of a classical string on a spacetime background with the metric $G^{\mu\nu}$ are governed by the Nambu-Goto action\footnote{The five-dimensional gravity solution has a 2-form field strength and a dilation field turned on, which could contribute to the string action. However, when this solution is lifted to ten dimensions on a 5-sphere, these fields become part of the metric and one can then use the Nambu-Goto action on the ten-dimensional background. It can be shown that the resulting equations of motion are satisfied by strings which are fixed at a single point on the 5-sphere, and this subset of string solutions can be obtained by considering the Nambu-Goto action on the five-dimensional background.}
\be
S=-T_0 \int d\sigma d\tau \sqrt{-g}\,,
\ee
where $(\sigma,\tau)$ are the worldsheet coordinates, $g_{ab}$ is the induced metric and $g=\det (g_{ab})$. The string tension $T_0$ is given by
\be
T_0=\fft{\sqrt{\lambda}}{2\pi \ell^2}\,.
\ee
For the map $X^{\mu}(\tau,\sigma)$ from the string worldsheet into spacetime,
\be
-g=(\dot X\cdot X^{\prime})^2-(X^{\prime})^2 (\dot X)^2\,,
\ee
where $\dot X=\partial_{\tau}X$, $X^{\prime}=\partial_{\sigma}X$ and the dot product of two vectors is given by $A\cdot B=A^{\mu}B^{\nu}G_{\mu\nu}$. We will study open strings on a background with the metric given by (\ref{metric}). We will consider strings which lie within a three-dimensional slice of the asymptotically AdS space in which $y=z=0$. Then $X(\sigma,\tau)$ is a map to $(t,\rho,x)$. Choosing a static gauge $\sigma=\rho$ and $\tau=t$, the string worldsheet is described by the function $x(t,\rho)$ and the equation of motion for the string is
\be\label{eom}
\fft{\partial}{\partial \rho}\left( f(\rho^2+q)\fft{x^{\prime}}{\sqrt{-g}}\right)-\fft{\rho^2+q}{f}\fft{\partial}{\partial t}\left( \fft{\dot x}{\sqrt{-g}}\right)=0\,,
\ee
where
\be\label{g}
-\fft{g}{\ell^4}=1-f^{-1}(\rho^2+q)\dot x^2+f(\rho^2+q)x^{\prime 2}\,.
\ee

The general expressions for the canonical momentum densities of the string are given by
\bea\label{momenta}
\pi_{\mu}^0 &=& -T_0 G_{\mu\nu} \fft{(\dot X\cdot X^{\prime})(X^{\nu})^{\prime}-(X^{\prime})^2 (\dot X^{\nu})}{\sqrt{-g}}\,,\nn\\
\pi_{\mu}^1 &=& -T_0 G_{\mu\nu} \fft{(\dot X\cdot X^{\prime})(\dot X^{\nu})-(\dot X)^2 (X^{\nu})^{\prime}}{\sqrt{-g}}\,.
\eea
The string tension $T_0=\sqrt{\lambda}/(2\pi\ell^2)$, where $\lambda$ is the 't Hooft coupling of the two field theories.

\subsection{Straight strings}

A steadily-moving straight string is given by the solution $x(t,\rho)=x_0+vt$, for which
\be
-g=1-f^{-1}(\rho^2+q)v^2\,.
\ee
If $\mu=0$, then $-g=1-v^2>0$ everywhere for $v<1$, which implies that all points along the string are moving at speeds less than the local speed of light. However, for $\mu\ne 0$, $-g$ is danger of vanishing at critical radii $\rho=\pm \rho_c$, where
\be
\rho_c^2 =\sqrt{\fft{\mu}{1-v^2}}-q\,.
\ee
If $\mu<q^2$, then the background is a wormhole and steadily-moving straight strings with speeds $|v|<v_{crit}$ are permissible, where
\be
v_{crit}=\sqrt{1-\fft{\mu}{q^2}}\,.
\ee
However, straight strings traveling at speeds $|v|\ge v_{crit}$ have a region for which $-g<0$ and the action, energy and momentum are complex. This is a signal that this portion of the string is traveling faster than the local speed of light and must be discarded. Thus, $v_{crit}$ corresponds to the maximum speed with which a quark in one of the dual field theories can move without undergoing energy loss.

In the next subsection, we will consider curved strings with $|v|>v_{crit}$ which satisfy $-g>0$ everywhere. Note that as $\mu\rightarrow q^2$ from below, $v_{crit}\rightarrow 0$. This means that for $\mu>q^2$, in which case horizons are present, steadily-moving straight strings are not permissible at any speed. This is a generic feature of the behavior of straight strings on black hole backgrounds, including those in gauged supergravity \cite{herzog,teaney,gubser,herzog2}. This implies that, for a finite-temperature field theory with a holographic description in terms of a black hole background, all moving quarks must lose energy.

\subsection{Curved strings}

We will now consider steadily-moving curved strings, which are described by solutions of the form $x(\rho,t)=x(\rho)+vt$. The term with the time derivative in (\ref{eom}) vanishes and we are left with
\be
\fft{d}{d\rho} \left( f(\rho^2+q)\fft{x^{\prime}}{\sqrt{-g/\ell^4}}\right) =0\,,
\ee
which can be integrated once to obtain
\be
f(\rho^2+q)\fft{x^{\prime}}{\sqrt{-g/\ell^4}}=Cv\,,
\ee
where $C$ is an integration constant. This constant determines the momentum current flowing along the string. From (\ref{momenta}), one finds
\be
\pi_t^1=\fft{\sqrt{\lambda}Cv^2}{2\pi}\,,\qquad \pi_x^1=-\fft{\sqrt{\lambda}Cv^2(\rho^2+q)^2}{2\pi\left[ (\rho^2+q)^2-\mu\right]}\,.
\ee
Curiously enough, $\pi_t^1$ is constant along the length of the string while $\pi_x^1$ is not.

Substituting $x^{\prime}$ into (\ref{g}) yields
\be
-\fft{g}{\ell^4}=\fft{(\rho^2+q)^2(1-v^2)-\mu}{(\rho^2+q)^2-\mu-C^2 v^2}\,.
\ee
For $|v|<v_{crit}$, $-g>0$ everywhere along the string provided that
\be
C^2<\fft{q^2-\mu}{v^2}\,.
\ee
Reality of $C$ implies that this condition cannot be satisfied for the nonsingular black hole, for which $\mu>q^2$.
For greater speeds, or for the regime of the nonsingular black hole, one must ensure that $-g$ remains positive everywhere by choosing $C$ such that the numerator and denominator of $-g$ change sign at the same point. This condition implies
\be\label{C}
C=\pm\sqrt{\fft{\mu}{1-v^2}}\,.
\ee
Then $-g/\ell^4=1-v^2>0$ provided that $|v|<1$. We also have
\be
x^{\prime}=\fft{\sqrt{\mu} v}{(\rho^2+q)^2-\mu}\,.
\ee
For the case of a wormhole ($\mu<q^2$), this can be integrated to yield the solution
\be\label{solution}
x(\rho,t)=x_0+vt +\fft{v}{2Q_-} \arctan \left(\fft{\rho}{Q_-}\right) -\fft{v}{2Q_+}\arctan\left( \fft{\rho}{Q_+}\right)\,,
\ee
where
\be\label{Q}
Q_{\pm}=\sqrt{q\pm \sqrt{\mu}}\,.
\ee
Note that we have taken the positive sign in (\ref{C}), which specifies that the forward endpoint of the string is the one located at $\rho>0$. This solution is shown in the left plot of Figure 1 for $\mu=1$, $q=2$, $v=0.1$ and $t=x_0=0$.
\begin{figure}[ht]
\begin{center}
$\begin{array}{c@{\hspace{.70in}}c}
\epsfxsize=2.6in \epsffile{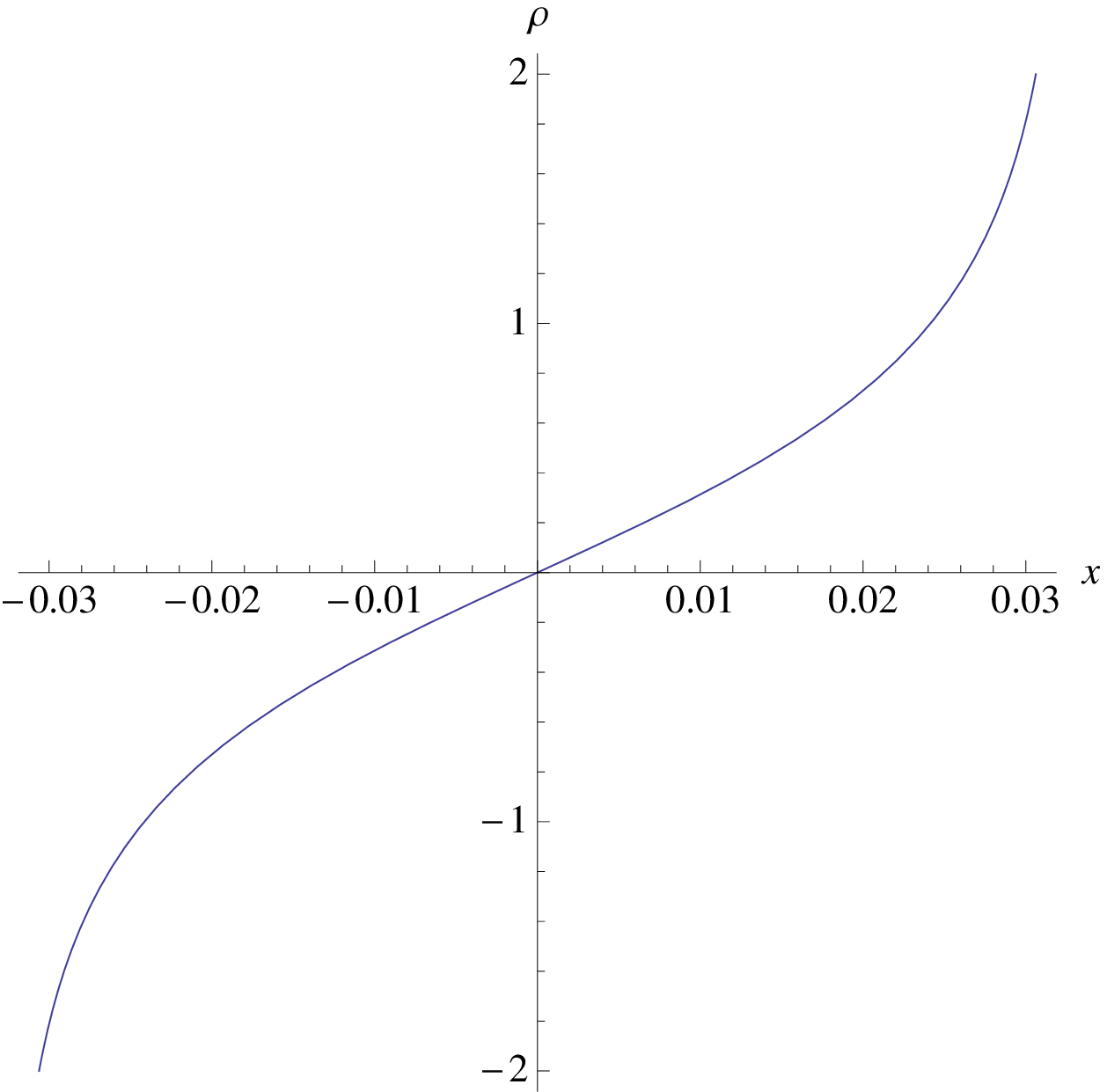} &
\epsfxsize=2.6in \epsffile{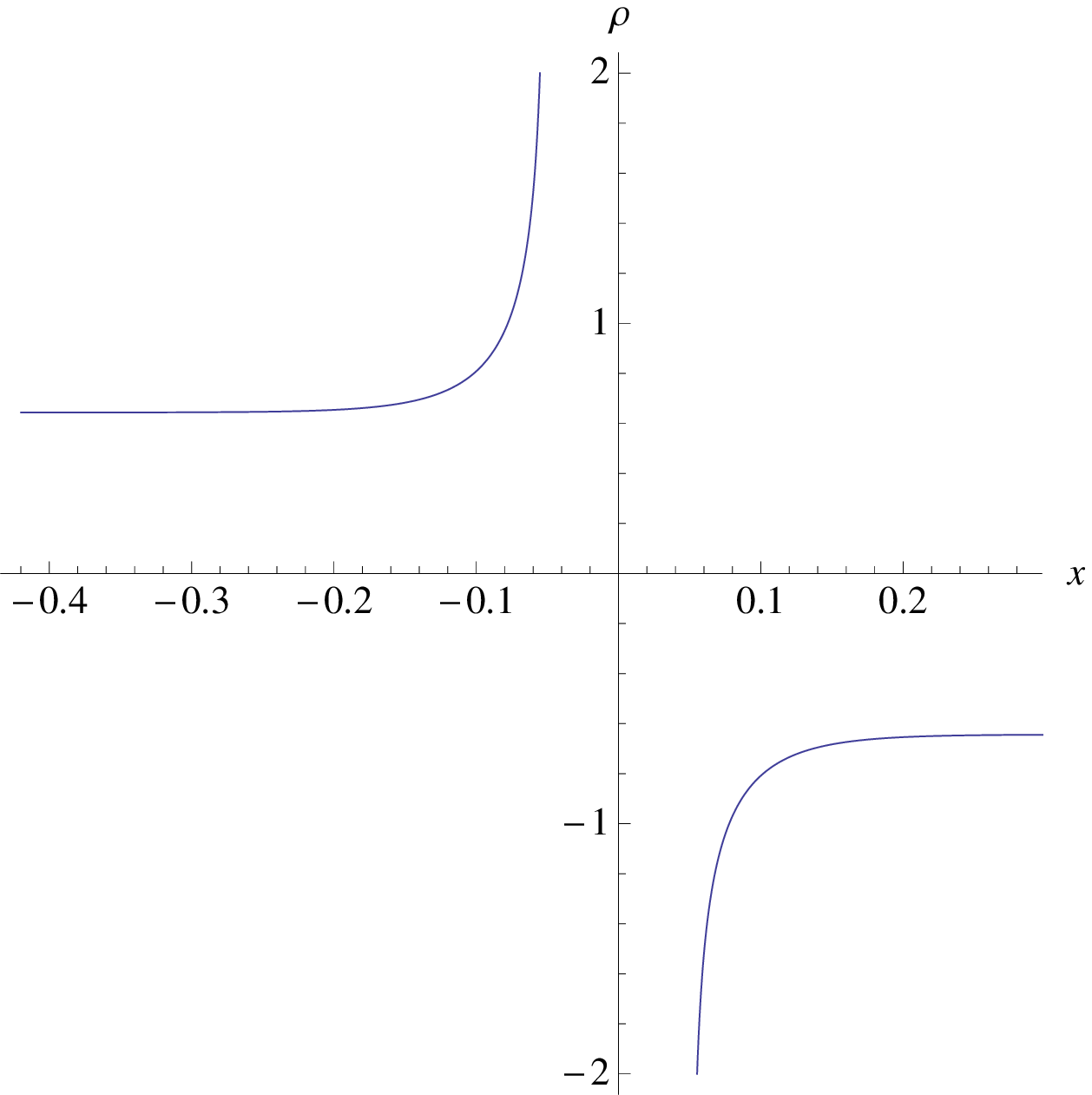}
\end{array}$
\end{center}
\caption[FIG. \arabic{figure}.]{\footnotesize{The left plot shows a curved string steadily moving in the positive $x$ direction in a wormhole background for $\mu=1$, $q=2$, $v=0.1$ and $t=x_0=0$. The right plot shows two disconnected strings moving in opposite directions in the black hole background for $\mu=2$, $q=1$, $v=0.1$ and $t=x_0=0$. The black hole horizons are located at $\rho\approx\pm 0.64$.}}
\end{figure}
For the case of a black hole ($\mu>q^2$), there are semi-infinite string configurations that occupy the regions outside of the horizons and are given by
\be
x(\rho,t)=x_0+vt +\fft{v}{4Q_-} \mbox{ln}\left(\fft{\rho+|Q_-|}{\rho-|Q_-|}\right) +\fft{v}{2Q_+}\arctan\left( \fft{\rho}{Q_+}\right)\,.
\ee
The analogous string solutions for AdS black holes in gauged supergravity were considered in \cite{herzog,teaney,gubser,herzog2}. Two such disconnected solutions moving in opposite directions along the $x$ axis are shown in the right plot of Figure 1 for $\mu=2$, $q=1$, $v=0.1$ and $t=x_0=0$. The rate at which energy flows along each of these strings is given by
\be
\pi_t^1=\pm \fft{v^2}{2\pi} \sqrt{\fft{\mu}{1-v^2}}\,.
\ee
For $\mu<q^2$, this corresponds to energy being transferred from a quark in one field theory to a quark in the other field theory. Similar string dynamics have been studied in \cite{gauss-bonnet} for wormhole backgrounds in Einstein-Gauss-Bonnet theory \cite{gb-wormhole1,gb-wormhole2} as well as a multi-boundary orbifold of AdS$_3$ \cite{orbifold}. In the case of $\mu>q^2$, energy is lost to the two surrounding plasmas. This is analogous to the energy loss that was studied using a curved string moving in an AdS black hole background in gauged supergravity \cite{herzog,teaney,gubser,herzog2}. However, in the present case the black hole has two horizons that are located at equal distance from $\rho=0$, which means that a quark in one field theory loses energy to both of the plasmas in equal amounts.

In the region that lies between the two horizons of the black hole, there is an infinitely extended string configuration given by
\be\label{behind-horizon}
x(\rho,t)=x_0+vt +\fft{v}{2|Q_-|} \mbox{arctanh} \left(\fft{\rho}{|Q_-|}\right) +\fft{v}{2Q_+}\arctan\left( \fft{\rho}{Q_+}\right)\,,
\ee
This configuration is shown in Figure 2 for $\mu=2$, $q=1$, $v=0.1$ and $t=x_0=0$.
\begin{figure}[ht]
   \epsfxsize=2.6in \centerline{\epsffile{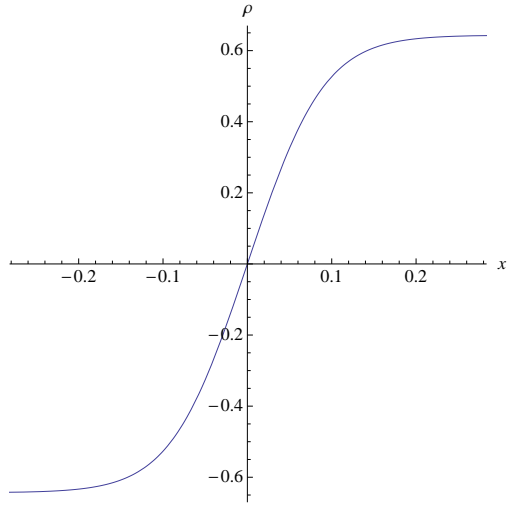}}
   \caption[FIG. \arabic{figure}.]{\footnotesize{A string configuration within the horizons of the black hole for $\mu=2$, $q=1$, $v=0.1$ and $t=x_0=0$.}}
\end{figure}
It is not currently well understood what role, if any, string configurations within black hole horizons play in the dual field theory. Despite the fact that the region between the horizons is not causally connected to either boundary, there have been indications that the dual field theories may still encode at least some of the physics associated with this region. A rather intriguing possibility is that the solution given by (\ref{behind-horizon}) corresponds to some sort of coupled collective excitations in the dual field theories.

\section{Quark-antiquark potential and screening length}

\subsection{Quark-antiquark potential}

For the nonsingular black hole background, one can have a single string that stretches from one of the boundaries to the black hole. From the perspective of the field theory living on that boundary, this is realized by a timelike Wilson-Polyakov loop having a nonvanishing expectation value $\left< P_i\right>$ or $\big< P_i^{\dagger}\big>$, depending on the orientation of the string. The index $i=1$ or $2$ denotes the boundary at which the string ends. For a string with both endpoints on the same boundary, the two-point correlation of Wilson-Polyakov loops $\big< P_i^{\dagger} P_i\big>$ can be computed from the proper area of the string worldsheet \cite{wilson1,wilson2}, from which the potential energy of a quark-antiquark pair can be read off. 

At zero temperature, $\left< P_i\right> =\big< P_i^{\dagger}\big> =0$. At this point, the theories are no longer decoupled and now interact with each other. This is embodied by the gravity background description in terms of a wormhole for which the two boundaries are in causal contact. Although the precise nature of the interaction between the two theories is not known, we propose that strings stretching 
between the two boundaries are the realizations of two-point Wilson-Polyakov loop correlators 
$\big< P_1^{\dagger} P_2\big>$ and $\big< P_2^{\dagger} P_1\big>$, and that these can also be computed from the proper area of the string worldsheet.

For the Euclideanized form of the metric (\ref{metric}), the Nambu-Goto action is given by
\be
S=T_0 \int d\sigma d\tau \sqrt{g}\,,
\ee
Using the static gauge $t=\tau$ and $x=\sigma$, this becomes
\be
S=2T_0 \Delta t \int_0^{L/2} dx \sqrt{f(\rho^2+q)+\rho^{\prime 2}}\,,
\ee
where $\Delta t$ is the time interval. Since the action does not depend on $x$ explicitly, the Beltrami identity yields
\be
\fft{f(\rho^2+q)}{\sqrt{f(\rho^2+q)+\rho^{\prime 2}}}=c\,.
\ee
The constant of motion $c$ is related to the minimal value of $\rho=\rho_0$ of the string, which occurs at $x=0$, by
\be
c=(\rho_0^2+q)^2-\mu\,.
\ee
$x$ as a function of $\rho$ is given by
\be
x=c \int \fft{d\rho}{\sqrt{\left( (\rho^2+q)^2-\mu\right) \left( (\rho^2+q)^2-\mu-c^2\right)}}\,,
\ee
where the lower bound of the integral is taken at $\rho=\rho_0$. For the case in which both endpoints of the string lie in the same asymptotic region, the integration constant $c$ is related to the distance $L$ between the quark and anti-quark by
\be
L=2c \int_{\rho_0}^{\infty} \fft{d\rho}{\sqrt{\left( (\rho^2+q)^2-\mu\right) \left( (\rho^2+q)^2-\mu-c^2\right)}}\,.
\ee
The energy of a static configuration is related to the action by $E=S/\Delta t$, which yields
\be
E=\fft{\sqrt{\lambda}}{\pi} \left[ \int_{\rho_0}^{\infty} d\rho \left( \sqrt{\fft{(\rho^2+q)^2-\mu}{(\rho^2+q)^2-\mu-c^2}}-1\right)-(\rho_0-\rho_h)\right]\,,
\ee
where we have regularized the energy by subtracting that of two disconnected straight strings which stretch from the horizon to infinity. For small $T$ and $\Phi$, the energy of a quark-antiquark pair is
\be
E=-\fft{4\pi^2 \sqrt{\lambda}}{\Gamma (\ft14)^4 L}\left[ 1+\fft{\Gamma(\ft14)^8}{240\pi^5}\left( -5(\ell^2\Phi L)^2+9\pi (\ell^2\Phi L)^2 (TL)^2+9\pi^3 (TL)^4\right) +\cdots\right]\,.
\ee
For the wormhole ($\mu<q^2$) for small $\mu$ and $q$, the energy of a quark-antiquark pair is
\be
E=-\fft{4\pi^2 \sqrt{\lambda}}{\Gamma (\ft14)^4 L}\left[ 1+\fft{\Gamma(\ft14)^8}{1920\pi^6} (-10\pi qL^2+9\mu L^4)
+\cdots\right]\,.
\ee

For the wormhole, we can also consider the case in which a string extends from one asymptotic region to the other. As depicted by the left plot of Figure 3, the distance between the endpoints along the $x$ direction is given by $L$, which is
\be
L=c \int_{-\infty}^{\infty} \fft{d\rho}{\sqrt{\left( (\rho^2+q)^2-\mu\right) \left( (\rho^2+q)^2-\mu-c^2\right)}}\,.
\ee
The energy of the string is
\be
E=\fft{\sqrt{\lambda}}{2\pi} \int_{-\infty}^{\infty} d\rho \left( \sqrt{\fft{(\rho^2+q)^2-\mu}{(\rho^2+q)^2-\mu-c^2}}-1\right)\,.
\ee
Note that we have subtracted the energy of a straight string so that this represents the interaction energy of a quark in one field theory with an anti-quark in the other as a function of their distance $L$. For small $L$, this is given by
\be
E= \fft12 k L^2 (1+aL^2+\cdots)\,,
\ee
where
\be
k=\fft{\sqrt{\lambda}}{2\pi^2} Q_+ Q_- (Q_++Q_-),\qquad a=\fft{Q_+^2+3Q_+Q_-+Q_-^2}{8\pi^2}\,,
\ee
and $Q_{\pm}$ are given by (\ref{Q}). Thus, when the quark and anti-quark are a small distance $L$ apart, they behave as though they are attached by a spring with an effective spring constant $k$, which is shown as a function of $\mu$ for different values of $q$ in the right plot of Figure 3. This type of phenomenon has previously been studied in \cite{gauss-bonnet} for wormholes in Einstein-Gauss-Bonnet theory \cite{gb-wormhole1,gb-wormhole2}, a multi-boundary orbifold of AdS$_3$ \cite{orbifold}, and the Maldacena-Maoz wormhole \cite{maoz}. Note that as one approaches the transition point between a wormhole and a nonsingular black hole ($\mu\rightarrow q^2$ from below), $k\rightarrow 0$. This is consistent with what one might naively expect, since this marks the transition to entangled states.
\begin{figure}[ht]
\begin{center}
$\begin{array}{c@{\hspace{0.9in}}c}
\epsfxsize=1.9in \epsffile{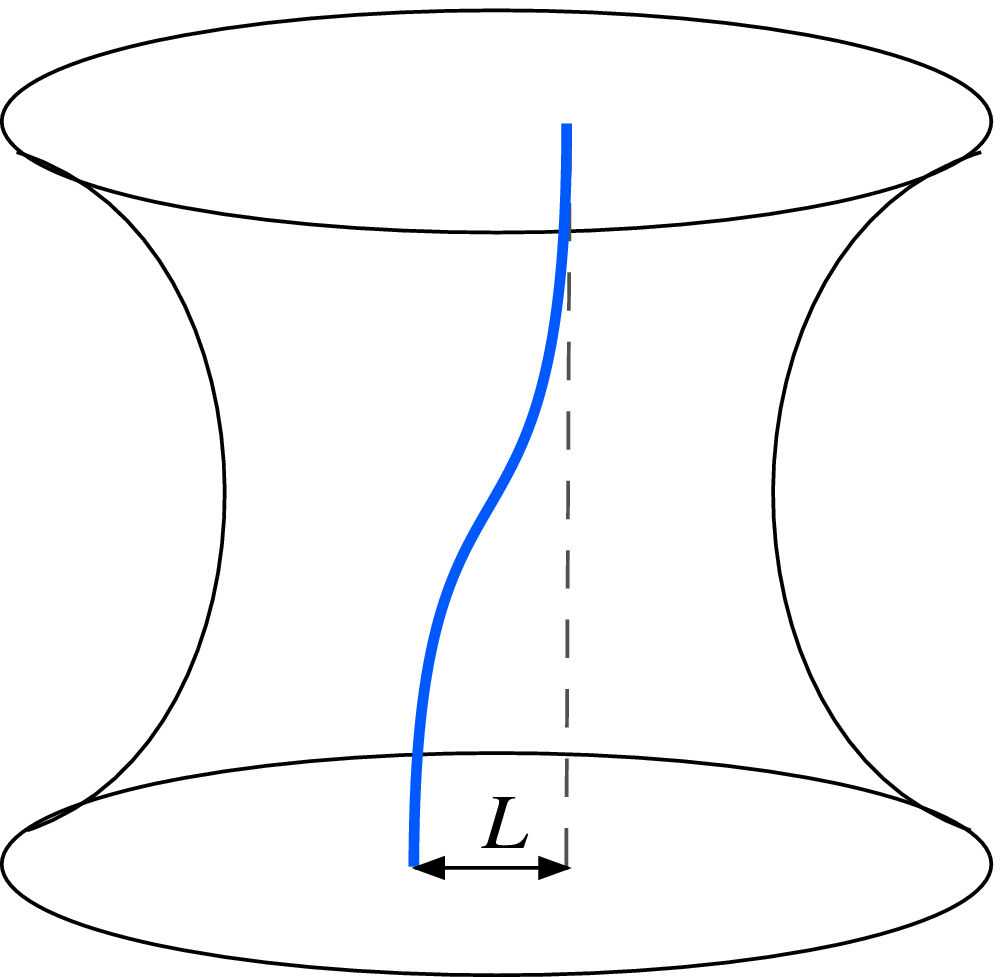} &
\epsfxsize=3.0in \epsffile{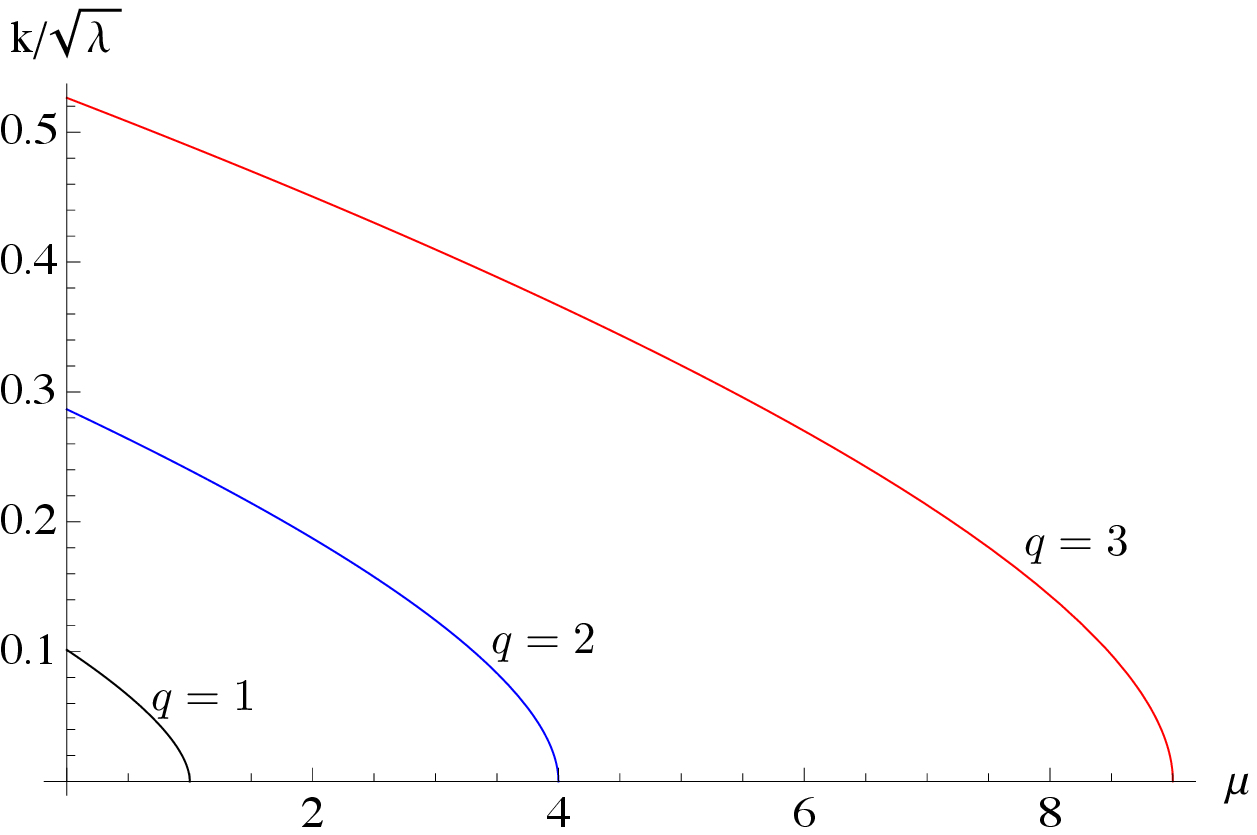}
\end{array}$
\end{center}
\caption[FIG. \arabic{figure}.]{\footnotesize{On the left is a schematic plot of a string stretching from one asymptotic region to the other. The right plot shows the effective spring constant $k$ connecting a quark and an anti-quark in different field theories as a function of $\mu$ for different values of $q$. We have taken $\lambda=1$.}}
\end{figure}

\subsection{Screening length}

The dependence of quark-antiquark screening length on temperature was studied in \cite{finiteT-wilson1,finiteT-wilson2} using a neutral AdS black hole in five-dimensional gauged supergravity. Its dependence on temperature and chemical potentials was studied in \cite{chemical-potential-wilson} using rotating nonextremal D3-branes, which correspond to a charged AdS black hole in five-dimensional gauged supergravity.

The screening length can be computed by comparing the energies of string configurations that share the same endpoints in the asymptotic region. As illustrated by the left plot of Figure 4, two points in the asymptotic region can either be the endpoints of a single curved string, or that of two disjoint strings extending straight down to the black hole horizon. If the distance $L$ between these two points is small, then the curved string remains far from the horizon and is energetically favorable. For large $L$, the curved string comes so close to the horizon that the disjoint straight strings are the energetically favorable configuration. The screening length $L_{screening}$ marks the distance at which this transition takes place. For $L>L_{screening}$, the interaction between the quark and the antiquark is completely screened by the plasma.

The right plot of Figure 4 shows the energy of a quark-antiquark pair as a function of $L$ for $T=0.2$ and $\Phi=1$. Recall that this energy has been regularized by subtracting the energy of the disjoint straight strings from that of the single curved string. Thus, the point at which the curve is at $E=0$ corresponds to the transition from the single curved string being the energetically favorable state (to the left of this point) to the disjoint straight strings being energetically favorable (to the right of this point). Notice that there is a portion of the curve which is always positive. This corresponds to a longer curved string that is always less energetically favorable than the aforementioned curved string as well as the disjoint straight strings. Also, the kink in the curve, corresponding to a maximum value that $L$ can take for the curved strings, occurs in the region for which the disjoint straight strings are energetically favorable. These non-energetically favorable string configurations correspond to quark-antiquark excitations that will eventually decay to the stable states.
\begin{figure}[ht]
\begin{center}
$\begin{array}{c@{\hspace{0.6in}}c}
\epsfxsize=2.2in \epsffile{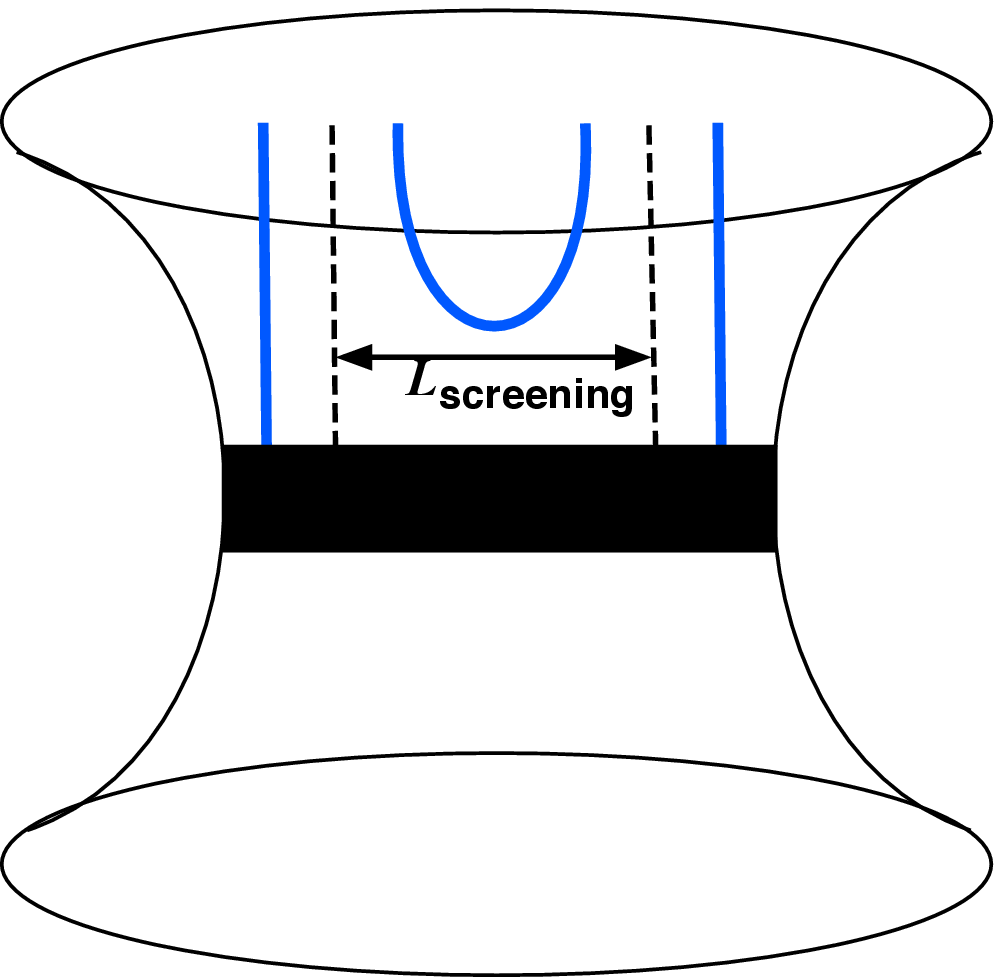} &
\epsfxsize=3.0in \epsffile{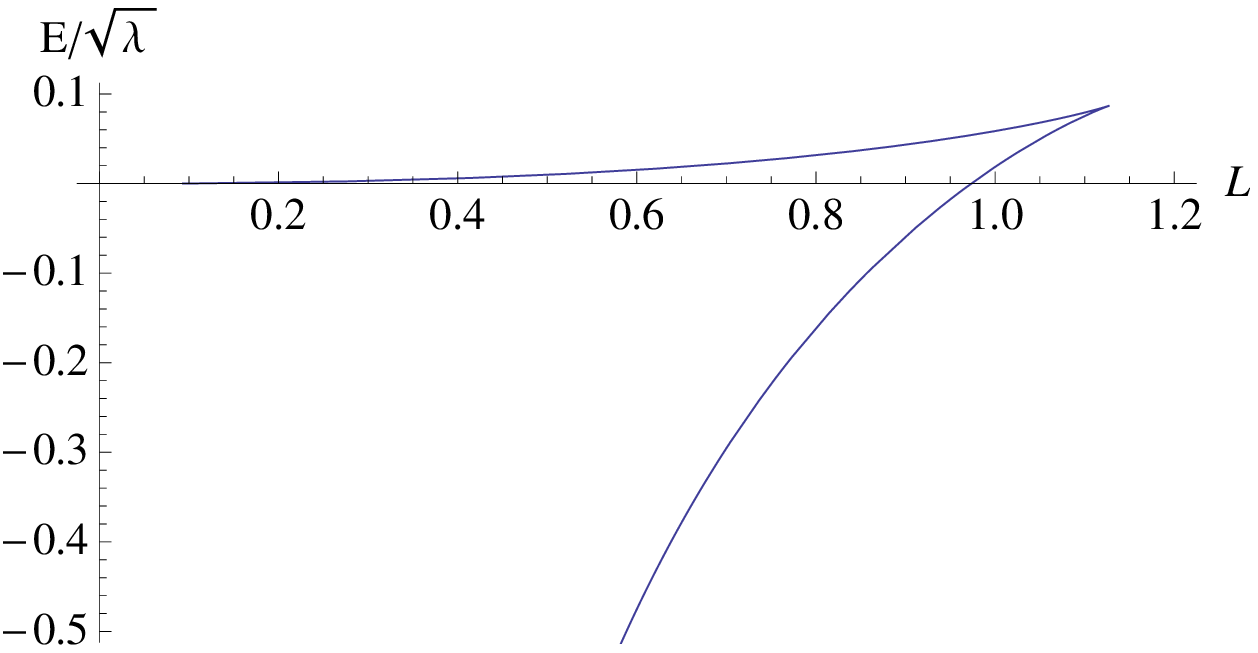}
\end{array}$
\end{center}
\caption[FIG. \arabic{figure}.]{\footnotesize{The left plot illustrates the interpretation of the screening length in terms of the energetics of string configurations on the nonsingular black hole background. The right plot shows the regularized energy versus $L$ for a quark-antiquark pair for $T=0.2$ and $\Phi=1$.}}
\end{figure}

For the nonsingular black hole ($\mu>q^2$, or equivalently $T>0$), we compute the screening length for a quark and an anti-quark as a function of temperature $T$ and chemical potential $\Phi$. The left plot in Figure 5 shows that the screening length monotonically decreases with temperature for fixed values of chemical potential. Indeed, it is certainly to be expected that the plasma blocks the interaction between the quark and antiquark at smaller distances at high temperature. Note that the screening length is independent of the chemical potential at high temperatures, as demonstrated by the convergence of the curves as $T$ increases in the left plot, as well as the flat line for $T=3$ in the right plot. Indeed, at high enough temperature,  adding more quark-antiquark pairs should not have a further effect on screening.
\begin{figure}[ht]
\begin{center}
$\begin{array}{c@{\hspace{.010in}}c}
\epsfxsize=2.95in \epsffile{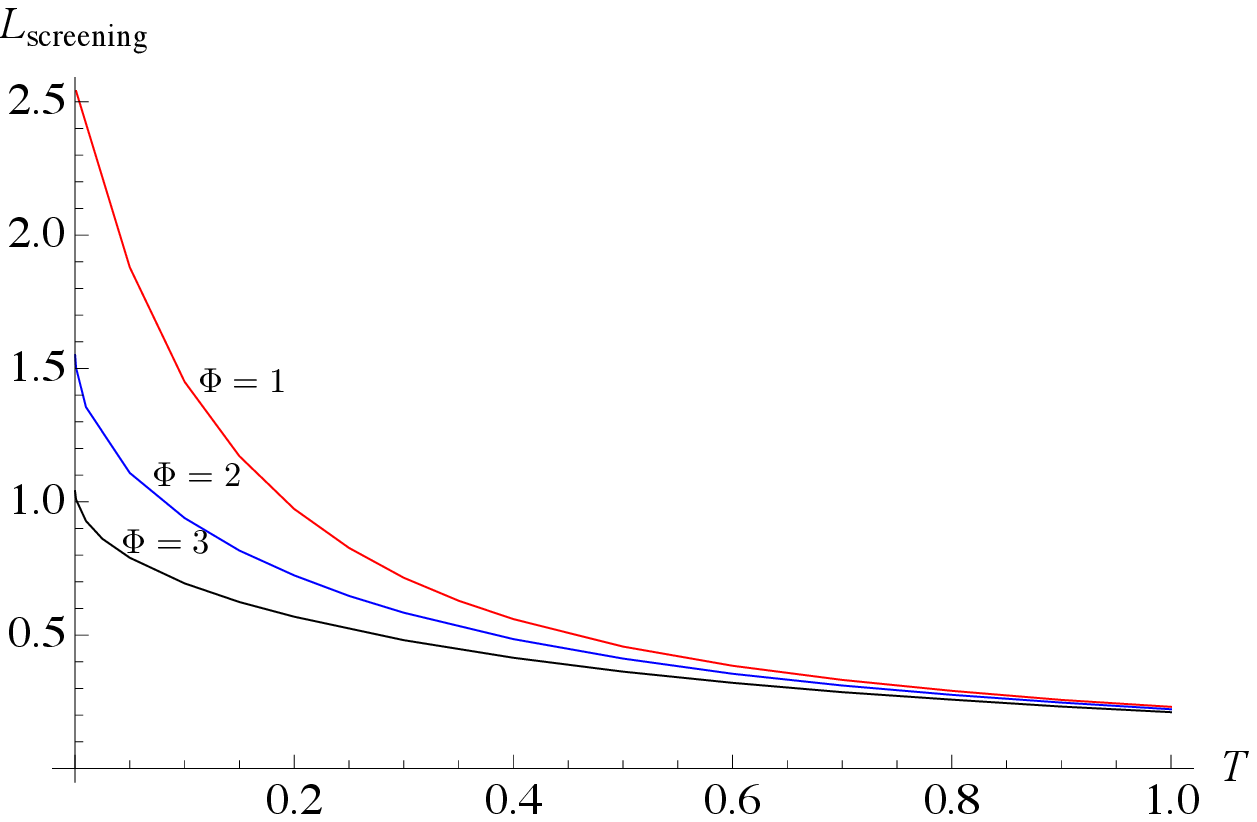} &
\epsfxsize=2.95in \epsffile{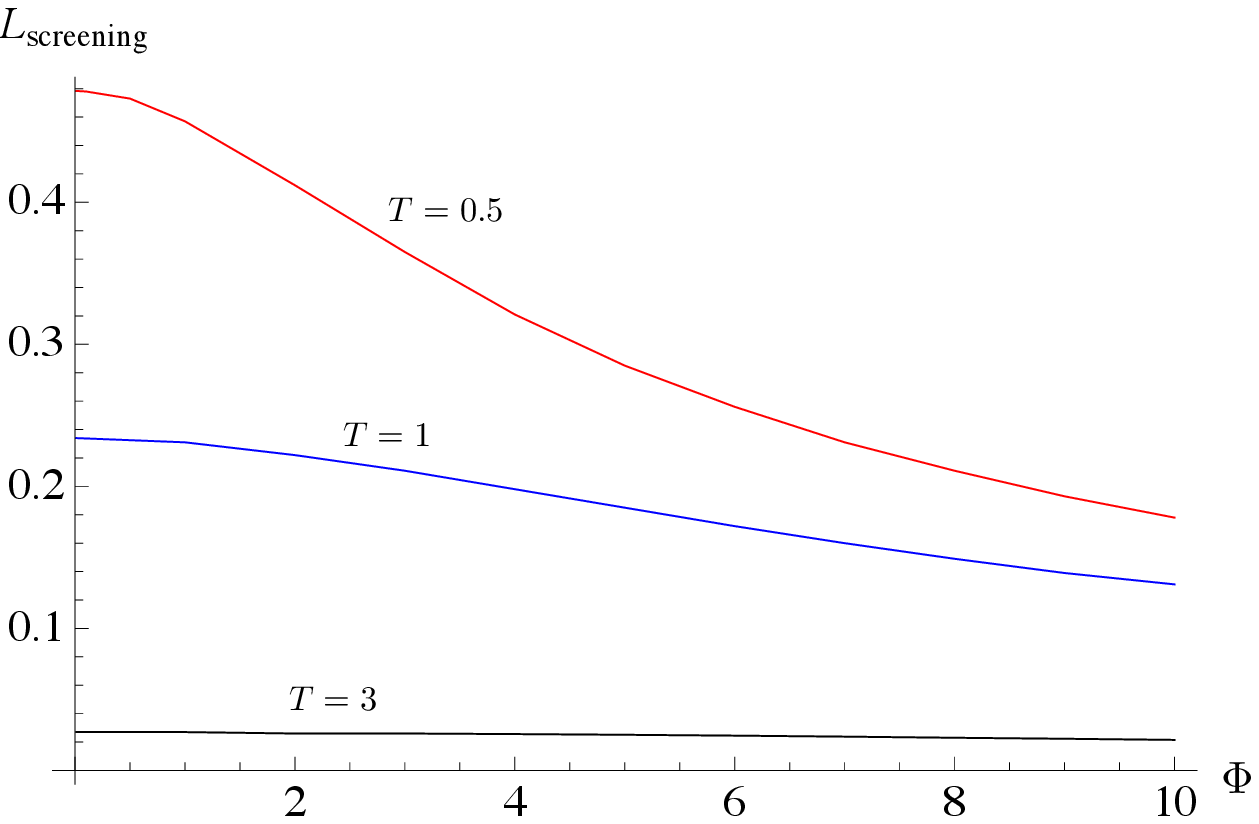}
\end{array}$
\end{center}
\caption[FIG. \arabic{figure}.]{\footnotesize{Quark-antiquark screening length as a function of temperature for various values of chemical potential (left) and as a function of chemical potential for various values of temperature (right).}}
\end{figure}

The right plot of Figure 5 also shows that the screening length monotonically decreases with chemical potential for fixed values of temperature. This is also to be expected, since a larger chemical potential implies that greater energy is for the production of additional quark-antiquark pairs to screen the interaction of the original quark and antiquark. Note that the screening length is not sensitive to variations of the chemical potential $\Phi$ for small $\Phi$, as shown by the flattening out of the curves in that regime. This could represent a type of saturation in the sense that such a large number of quark-antiquark pairs have already been produced, since such little energy is needed to produce each pair, that introducing additional pairs into the system would not have a further effect on screening.
\begin{figure}[ht]
\begin{center}
$\begin{array}{c@{\hspace{.30in}}c@{\hspace{.30in}}c}
\epsfxsize=1.7in \epsffile{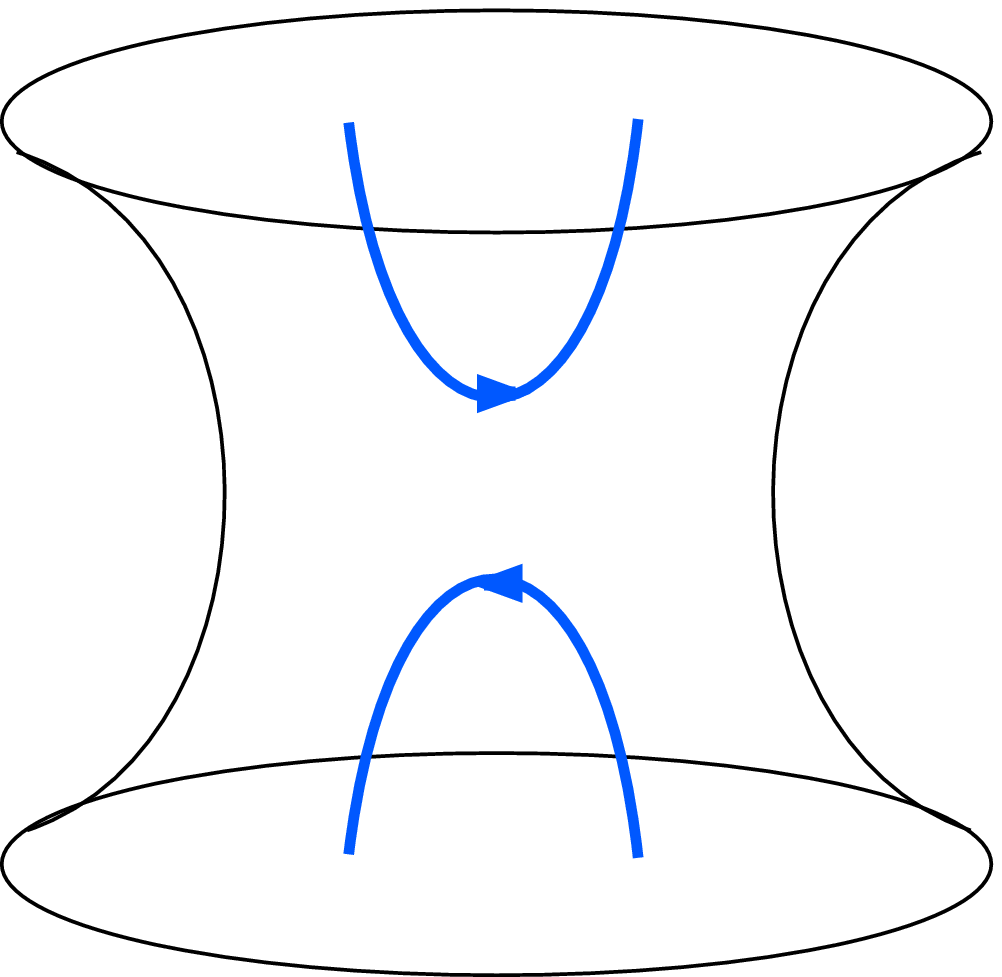} &
\epsfxsize=1.7in \epsffile{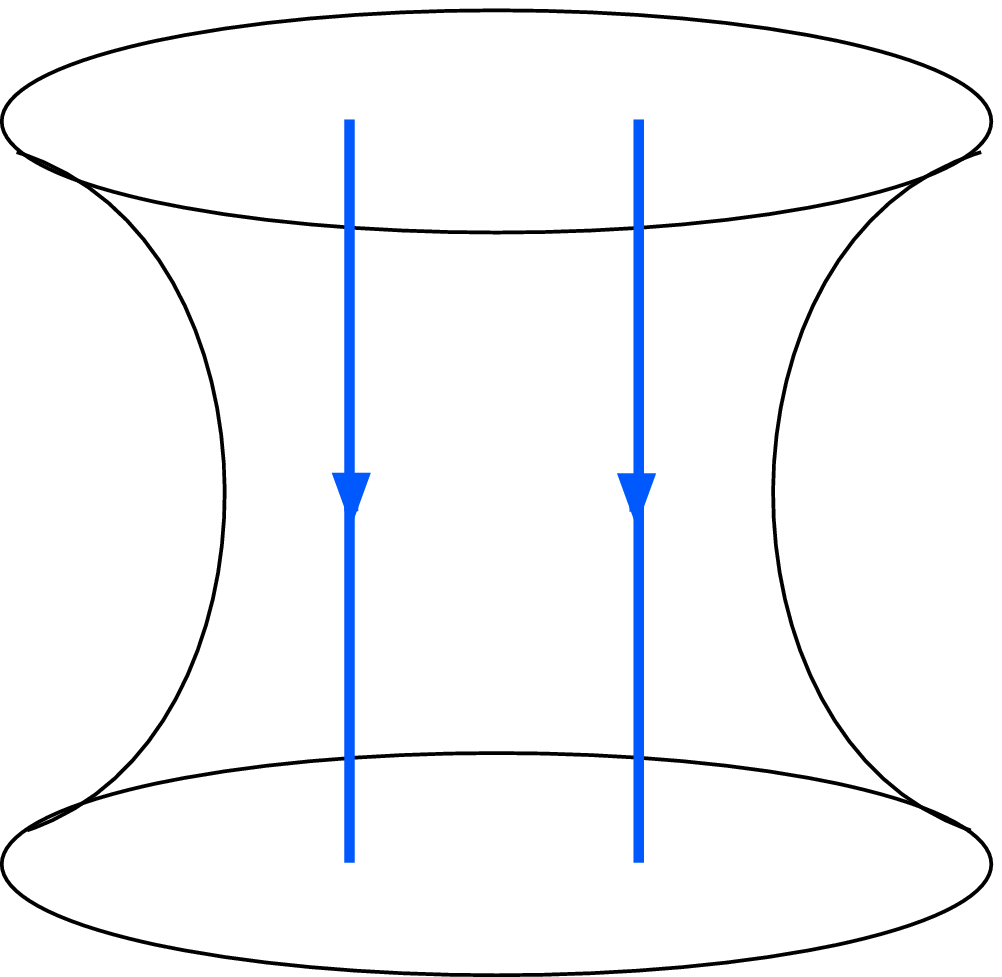} &
\epsfxsize=1.7in \epsffile{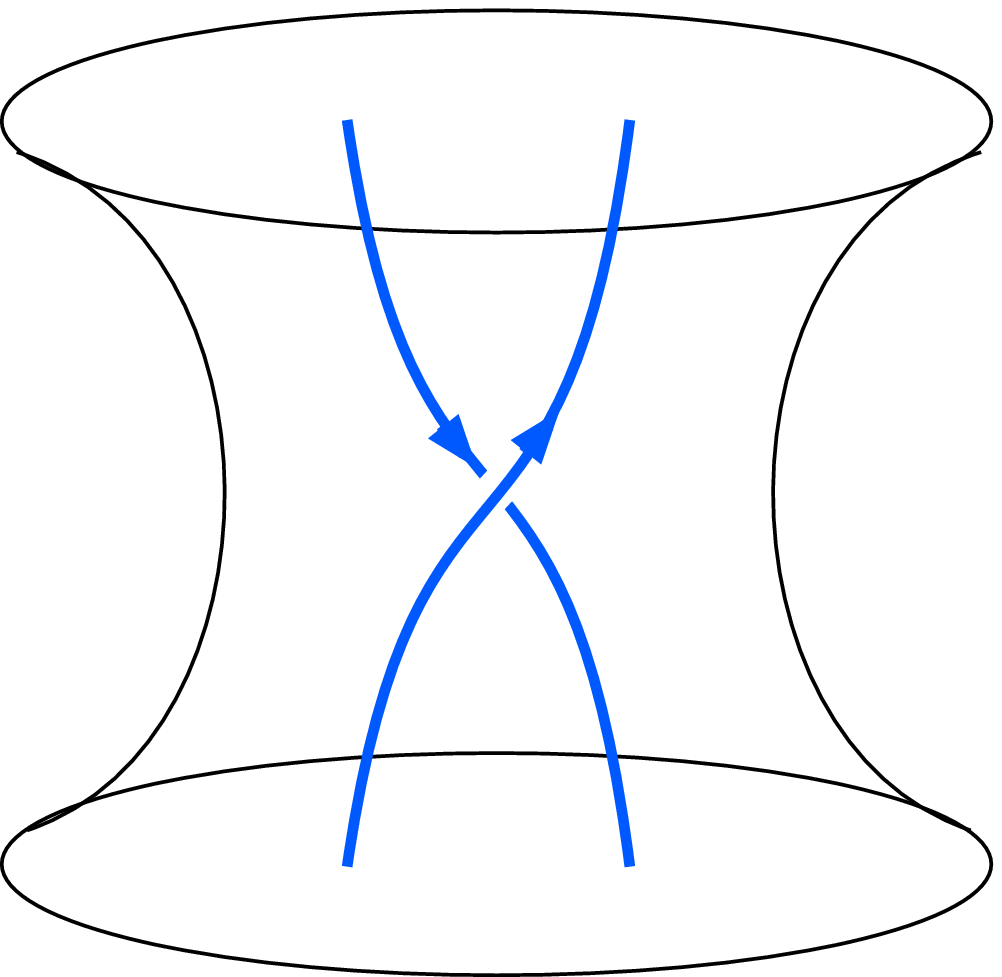}
\end{array}$
\end{center}
\caption[FIG. \arabic{figure}.]{\footnotesize{Schematic plots of three different string configurations on the wormhole background that have common locations for their endpoints in the asymptotic regions.}}
\end{figure}

Figure 6 illustrates some configurations involving a pair of strings that can exist on a wormhole background, for which some interesting transitions can take place. We will first consider the simplest situation for which both pairs  of endpoints on each side of the wormhole are a distance $L$ apart and are positioned in a symmetric manner. Although each configuration can have common locations for their endpoints, in order to do a comparison of their energies, the orientation of some of the strings must sometimes be reversed. As it stands, the U-shaped strings shown on the left in Figure 6 and the crossed strings on the right both represent a quark-antiquark pair in each field theory, whereas the straight strings in the center represent two quarks in one field theory and, by fiat, two antiquarks in the other.

One can readily surmise that the two U-shaped strings are energetically favorable for small $L$ whereas, upon changing the orientation accordingly, the two straight strings are energetically favorable for large distance. Indeed, this is qualitatively like two copies of the screening effect that takes place for the case of a black hole. However, this screening exists at zero temperature and is brought about by the interaction of a quark in one field theory with an antiquark in the other one, rather than by the effects of a plasma.
\begin{figure}[ht]
\begin{center}
$\begin{array}{c@{\hspace{.30in}}c}
\epsfxsize=2.8in \epsffile{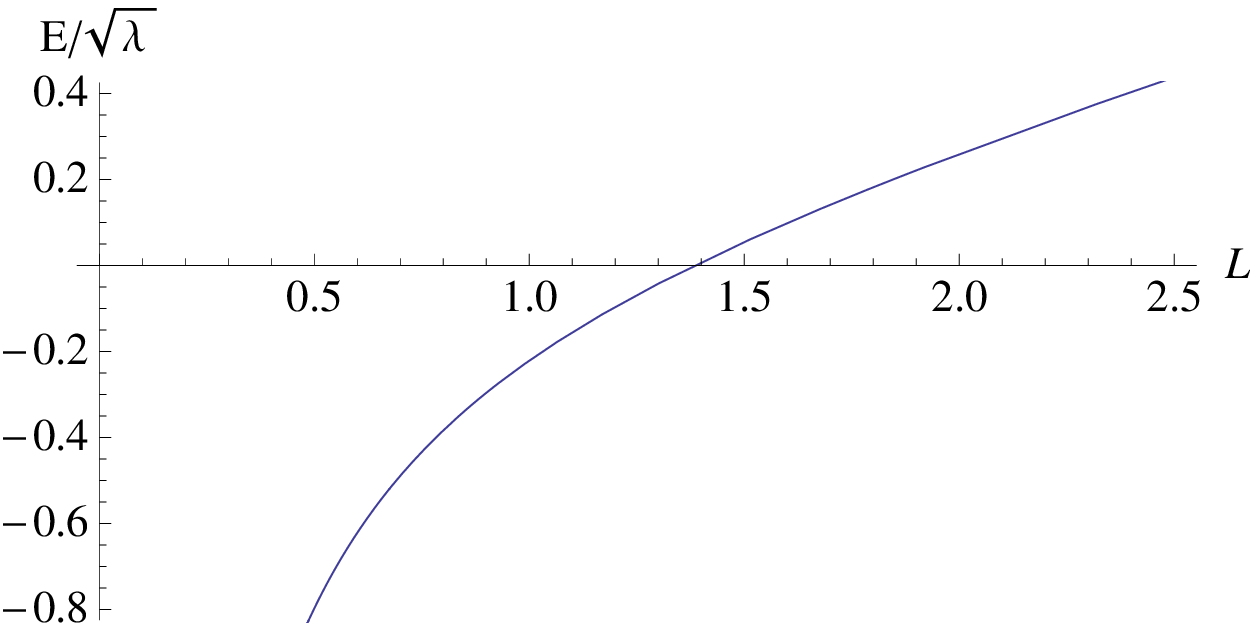} &
\epsfxsize=2.8in \epsffile{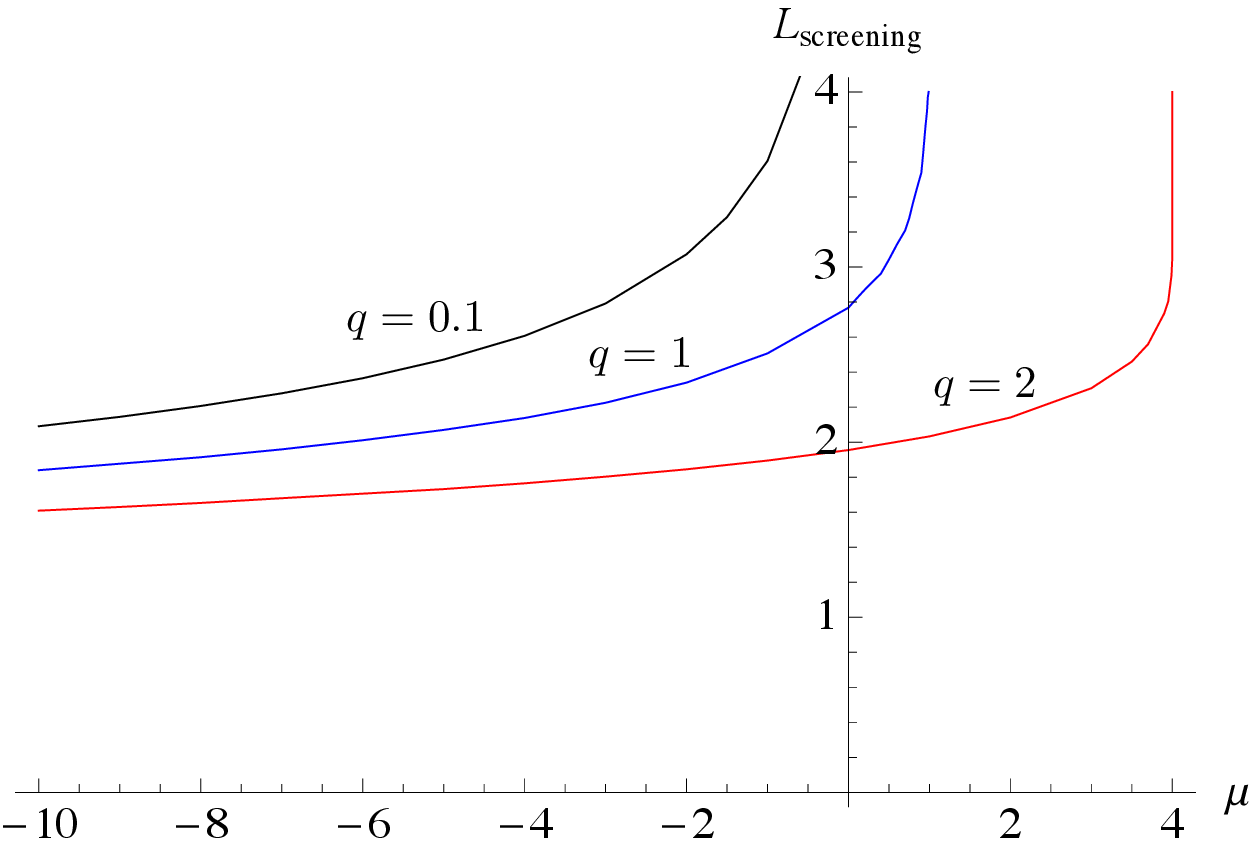}
\end{array}$
\end{center}
\caption[FIG. \arabic{figure}.]{\footnotesize{The left plot shows the regularized energy versus $L$ for a quark-antiquark pair in each field theory with $\mu=0$ and $q=1$. The right plot shows the screening length as a function of $\mu$ for various values of $q$.}}
\end{figure}

The energy of the U-shaped strings as well as the crossed strings can be regularized by subtracting that of the two straight strings. The left plot in Figure 7 shows the regularized energy for the U-shaped strings for $\mu=0$ and $q=1$, which illustrates our above expectation. The value of $L$ for which this curve passes through zero energy represents a screening length in the sense that the quark-antiquark pair in each field theory do not interact with each other for distances greater than that. It turns out that the crossed string configuration shown on the right in Figure 6 always has greater energy than both the U-shaped strings as well as the straight strings. Thus, at least for this symmetric positioning of endpoints, the crossed string configuration is never energetically favorable.
\begin{figure}[ht]
   \epsfxsize=2.2in \centerline{\epsffile{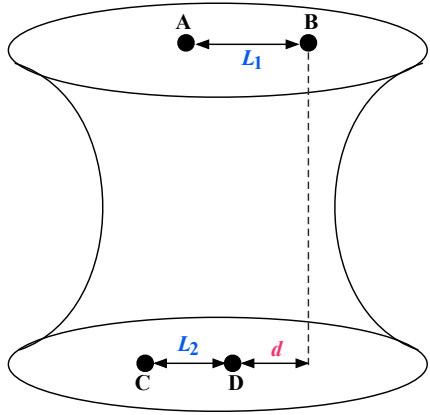}}
   \caption[FIG. \arabic{figure}.]{\footnotesize{Points A, B, C and D represent string endpoints in the asymptotic regions. $L_1$ and $L_2$ are the distances of each pair of endpoints in each asymptotic region, and $d$ is the relative displacement of the pair of endpoints in one asymptotic region from that in the other.}}
\end{figure}

The situation becomes more intricate for less symmetric positioning of the endpoints, as shown in Figure 8. Points A and B on one side of the wormhole are a distance $L_1$ apart, while points C and D on the other side are a distance $L_2$ apart. Also, points C and D can be shifted a distance $d$ relative to the other pair of points. We will restrict ourselves to string configurations that lie entirely in the $x-r$ plane at $y=z=0$.
The configuration for which one string connects points A and B and the other string connects points C and D can be denoted as the AB-CD configuration, which we will refer to as the U-shaped string configuration. In this vein, the AD-BC configuration can be referred to as the crossed string configuration. In the limit of $L_1=L_2$ and $d=0$, the AC-BD configuration reduces to that of two straight strings.

\begin{figure}[ht]
   \epsfxsize=2.8in \centerline{\epsffile{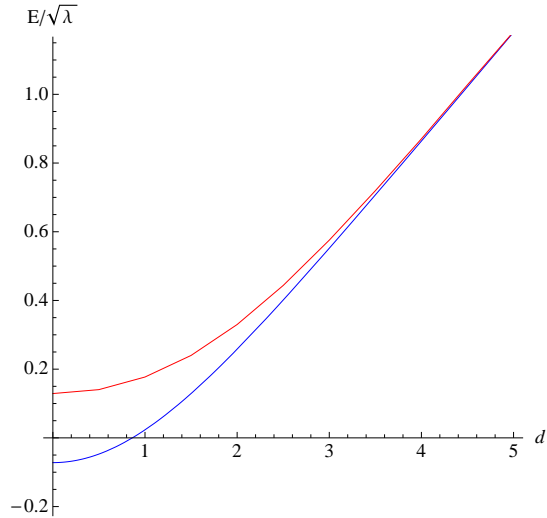}}
   \caption[FIG. \arabic{figure}.]{\footnotesize{The energies of the AC-BD configuration (blue curve) and the crossed string configuration (red curve) for $\mu=0$, $q=1$ and $L_1=L_2=1.5$. The energies have been regularized by subtracting that of the U-shaped string configuration. The AC-BD configuration is energetically favorable for $d<0.862$, while the U-shaped string configuration is energetically favorable for $d>0.862$.}}
\end{figure}
Let us first consider the situation for which $\mu=0$, $q=1$, $L_1=L_2$ is fixed and $d$ varies, in which case the energy of the U-shaped string configuration can be held constant. Therefore, the most convenient regularization scheme is to subtract this energy from that of the other two types of configurations in order to obtain their regularized energy. Note that this is a consistent regularization scheme despite the fact that the crossed strings have a different orientation from the U-shaped strings. The left plot of Figure 7 indicates that, for $\mu=0$ and $q=1$, the screening length for a quark-antiquark pair in each field theory is approximately $L_{screening}=1.4$. We will take $L_1=L_2=1.5$ so that the energetically favorable configuration for $d=0$ is the AC-BD configuration, which in this limit consists of two straight strings. The regularized energy of the string configurations as a function of $d$ is shown in Figure 9. From this, we can see that the AC-BD configuration is energetically favorable for $d<0.862$ and the U-shaped string configuration is energetically favorable for $d>0.862$. In other words, the screening length increases with $d$, so that $L_{screening}=1.5$ for $d=0.862$. On the other hand, there are no transitions involving the crossed string configuration, since it is the one that has the highest energy for all $d$, although its energy asymptotically approaches that of the AC-BD configuration for large $d$ as to be expected.

\begin{figure}[ht]
   \epsfxsize=3.0in \centerline{\epsffile{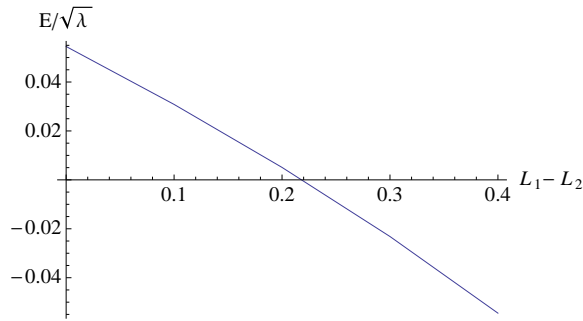}}
   \caption[FIG. \arabic{figure}.]{\footnotesize{The energy of the U-shaped string configuration regularized with respect to the AC-BD configuration as a function of $L_1-L_2$. For $L_1-L_2<0.22$ the AC-BD configuration is energetically favorable, while for $L_1-L_2>0.22$ the two U-shaped strings are the energetically favorable configuration.}}
\end{figure}
Now consider the situation for which $\mu=0$, $q=1$, $d=0$, $L_1=1.5$ and $L_2<L_1$. The comparison between only the U-shaped string configuration and the AC-BD configuration are presented, since the crossed string configuration is never energetically favorable. Since none of these string configurations have fixed energy as $L_2$ varies, we can just as well choose the regularization scheme of subtracting the energy of the AC-BD configuration from that of the U-shaped string configuration. This regularized energy as a function of $L_2-L_1$ is shown in Figure 10, from which we see that the AC-BD configuration is energetically favorable for $L_1-L_2<0.22$, while the U-shaped string configuration is energetically favorable for $0.22<L_1-L_2<1.5$. Despite the fact that the quark and antiquark in one field theory remain relatively far apart, a transition can be brought about so that their interaction is no longer screened by moving the quark and antiquark in the other field theory closer together.

\section{Conclusions}

We have investigated the behavior of open strings on five-dimensional wormhole and black hole backgrounds in five-dimensional gauged supergravity in the $f(R)$ frame. These backgrounds have two asymptotic AdS regions and are completely free of curvature singularities. They are conformally related to AdS black holes in the STU model with two equal charges. Although the backgrounds do not contain curvature singularities, they are supported by a delta-function source for a scalar field. The $f(R)$ frame corresponds to the dual frame, or effective string frame, which implies that classical string probes are not sensitive to this singularity. We have considered an effective string theory on these backgrounds and have made the working assumption that the AdS/CFT can be applied in this scenario. Then the wormhole background describes a phase in which the degrees of freedom on different boundaries interact with each other, whereas the nonsingular black hole describes an entangled state in two identical conformal field theories.

Studying the dynamics of open strings on these backgrounds enables us to extract some of the features of the quarks and anti-quarks that live in the field theories. For the phase in which there is a coupling between the degrees of freedom on different boundaries, there is a maximum speed with which the quarks can move without losing energy. If a quark moves faster than this, then energy will be transferred to a quark in the other field theory such that the two quarks start moving together. For the case of entangled states, a moving quark loses energy to the two surrounding plasmas in equal amounts, and we have computed the rate at which this rather unusual energy transfer process occurs.

We have also studied the potential energy between static quarks and anti-quarks. For small separation, a quark and an anti-quark exhibit Coulomb interaction if they are associated with the same boundary. On the other hand, in the interacting phase, a quark associated with one boundary exhibits spring-like confinement with an anti-quark associated with the other boundary. The dependence of the effective spring constant on the parameters of the wormhole background is presented. As might be expected, as one changes the parameters in such a way that the transition to entangled states is approached, the effective spring constant monotonically decreases to zero.

The screening length between quarks and anti-quarks have been investigated. For the case of entangled states, we have studied the dependence of the screening length on temperature and chemical potential and find that it monotonically decreases with each of these quantities. However, for either high temperature or small chemical potential, the screening length is not very sensitive to changes in the chemical potential, which represents a sort of saturation in the sense that introducing additional quark-antiquark pairs into the system would not have a further effect on screening.

For the phase in which degrees of freedom on different boundaries are interacting, there is an interesting transition that can take place involving a quadruplet made up of one quark-antiquark pair in each field theory. On one side of the transition the quarks interact with the antiquarks that are associated with the same boundary and are screened from antiquarks associated with the other boundary, whereas on the other side of the transition it is the reverse. We consider various ways in which the quarks and anti-quarks can be positioned so as to bring about this transition. It could also be interesting to study analogous transitions that can take place involving screening and two heavy-light mesons in a single field theory that has flavor.

\section*{Acknowledgments}

We are grateful to James Liu for helpful conversations. J.F.V.P. is grateful to Beijing Normal University for hospitality during the initial stages of this work. The work of H.L. was supported in part by the NSFC grants 11175269 and 11235003. The work of J.F.V.P. was supported in part by NSF grant PHY-0969482.

\appendix

\section{General black holes and thermodynamics}

The Lagrangian (\ref{d5suplag1}) for a five-dimensional theory can be generalized to arbitrary dimensionality.  In particular, it can be shown that the four-dimensional theory can be embedded within a supergravity theory. While the analogous theories with $D\ge 6$ cannot be embedded in supergravity, it was shown in \cite{Liu:2012jra,Liu:2011ve} that the theory with $A=0$ can be pseudo-supersymmetrized in that the full theory is invariant under the pseudo-supersymmetric transformation rules up to and including the quadratic order in fermionic fields. In fact, that was how the scalar potential was obtained.

We shall first study the theory in the Einstein frame and discuss the charged black hole solution obtained in \cite{susyfr}.  We analyse the global structure and obtain all the thermodynamical quantities.  We then analyse the solutions in the $f(R)$ frame and show that wormholes emerge.

\subsection{Einstein frame}

Let us consider the Lagrangian
\begin{equation}
e^{-1} {\cal L} = R - \ft12 (\partial\phi)^2  -\ft14 e^{2(D-3)\alpha \phi} F_\2^2- \ft14
e^{-2(D-1)\alpha \phi} {\cal F}_\2^2 - V(\phi)\,,\label{gaugegenlag}
\end{equation}
where the scalar potential is given by
\begin{equation}
V(\phi)=- (D-1)g^2\Big((D-3) e^{-\sqrt{\fft{2}{(D-1)(D-2)}} \,\phi}+
e^{\fft{\sqrt2\,(D-3)}{\sqrt{(D-1)(D-2)}}
\,\phi}\Big)\,.\label{scalarpot}
\end{equation}
The theory admits the following charged black hole solution \cite{susyfr}:
\begin{eqnarray}
ds_D^2&=&-{\cal H}^{-\fft{D-3}{D-2}} H^{-\fft{D-1}{D-2}}\, h\, dt^2
+ {\cal H}^{\fft{1}{D-2}} H^{\fft{D-1}{(D-2)(D-3)}}
\Big(\fft{dr^2}{h} + r^2 d\Omega_{D-2,\ep}^2\Big)\,,\cr 
{\cal F}_\2&=& \sqrt{\ep}\coth(\sqrt{\ep}\,\tilde\delta)\,dt\wedge
d{\cal H}^{-1}\,,\quad F_\2=\sqrt{\fft{D-1}{D-3}}\,\sqrt{\ep}\,
\coth(\sqrt{\ep}\,\delta)\,dt\wedge dH^{-1}\,,\cr 
h &=& \ep -\fft{\mu}{r^{D-3}} + g^2 r^2 {\cal H}
H^{\fft{D-1}{D-3}}\,,\qquad e^{\phi} = \Big(\fft{\cal
H}{H}\Big)^{\sqrt{\fft{D-1}{2(D-2)}}}\,,\cr 
{\cal H} &=& 1 +
\fft{\mu\sinh^2(\sqrt{\ep}\,\tilde\delta)}{\ep\,r^{D-3}}\,,\qquad H=1 +
\fft{\mu\sinh^2(\sqrt{\ep}\,\delta)}{\ep\,r^{D-3}}\,.\label{cbh0}
\end{eqnarray}
It is more convenient to use the parametrization
\begin{equation}
\tilde q=\mu \ep^{-1} \sin^2(\sqrt{ep} \tilde \delta)\,,\qquad
q=\mu \ep^{-1} \sin^2(\sqrt{\ep}\delta)\,.
\end{equation}
Then the solution becomes
\begin{eqnarray}
ds_D^2&=&-{\cal H}^{-\fft{D-3}{D-2}} H^{-\fft{D-1}{D-2}}\, h\, dt^2
+ {\cal H}^{\fft{1}{D-2}} H^{\fft{D-1}{(D-2)(D-3)}}
\Big(\fft{dr^2}{h} + r^2 d\Omega_{D-2,\ep}^2\Big)\,,\cr 
{\cal F}_\2&=& \sqrt{\ep + \mu \tilde q^{-1}}\,dt\wedge
d{\cal H}^{-1}\,,\quad F_\2=\sqrt{\fft{D-1}{D-3}}\,\sqrt{\ep + \mu q^{-1}}\,dt\wedge dH^{-1}\,,\cr 
h &=& \ep -\fft{\mu}{r^{D-3}} + g^2 r^2 {\cal H}
H^{\fft{D-1}{D-3}}\,,\qquad e^{\phi} = \Big(\fft{\cal
H}{H}\Big)^{\sqrt{\fft{D-1}{2(D-2)}}}\,,\cr 
{\cal H} &=& 1 +
\fft{\tilde q}{r^{D-3}}\,,\qquad H=1 +
\fft{q}{r^{D-3}}\,.\label{cbh1}
\end{eqnarray}
The horizon is located at the largest positive root of $h$, namely $h(r_0)=0$. It is convenient to solve for $\mu$ instead:
\begin{equation}
\mu= r_0^{D-3} \Big(\ep + g^2 r_0^2 {\cal H}_0 H_0^{\fft{D-1}{D-3}}\Big)\,,
\end{equation}
where ${\cal H}_0={\cal H}(r_0)$ and $H_0=H(r_0)$.
Using standard techniques, we find the following thermodynamical quantities:
\begin{eqnarray}
&&T=\fft{h'(r_0)}{4\pi {\cal H}_0^\fft12 H_0^{\fft{D-2}{2(D-3)}}}\,,\qquad
S=\fft{\omega}{4} {\cal H}_0^{\fft12} H_0^{\fft{D-1}{2(D-3)}} r_0^{D-2}\,,\cr
&&\tilde\Phi=\sqrt{\ep + \mu \tilde q^{-1}} ({\cal H}_0^{-1}-1)\,,\qquad
\tilde Q=\fft{(D-3)\omega}{16\pi}\, \tilde q \sqrt{\ep + \mu \tilde q^{-1}}\,,\\
&&\Phi=\sqrt{\fft{D-1}{D-3}}\,\sqrt{\ep + \mu q^{-1}} (H_0^{-1}-1)\,,\quad
 Q=\fft{\sqrt{(D-1)(D-3)}\,\omega}{16\pi}\, q \sqrt{\ep + \mu q^{-1}}\,,\cr
&&M=\fft{\omega}{16\pi}\Big[ \Big(\tilde q (D-3) + q (D-1) + (D-2) r_0^{D-3}\Big) \ep + (D-2) g^2 r_0^{D-1} {\cal H}_0 H_0^{\fft{D-1}{D-3}}\Big],\nn
\end{eqnarray}
where $\omega$ is the volume of $d\Omega_{D-2,\ep}^2$.  Note that the above quantities satisfy the first law of thermodynamics
\begin{equation}
dM=TdS + \tilde \Phi d\tilde Q + \Phi dQ\,.
\end{equation}
The mass can also be calculated directly and is given by
\begin{equation}
M=\fft{\omega}{16\pi}\Big( (D-2)\mu + (D-1) \ep \tilde q + (D-3) \ep q\Big)\,.
\end{equation}
For $\ep=1$ and $\tilde q q\ne 0$, the solution has a BPS-like limit of $\mu=0$, for which we have
\begin{equation}
M=\tilde Q + \sqrt{\fft{D-1}{D-3}}\, Q\,.\label{bps}
\end{equation}
The BPS-like condition (\ref{bps}) becomes precisely the BPS condition in four and five dimensions, for which
the theory can be embedded within supergravity. There is also an extremal limit of $\mu=\mu_0$, for which the near-horizon geometry is AdS$_2\times \Omega^{D-2,\ep}$, where $\Omega^{D-2,\ep}=S^{D-2}$, $T^{D-2}$ or $H^{D-2}$ for $\ep=1$, $0$ or $-1$, respectively.  The results in this section have been generalized to a wider class of AdS black holes \cite{Lu:2013eoa}.

\subsection{The $f(R)$ frame}

We now consider the $f(R)$ frame, namely
\begin{equation}
g_{\mu\nu} \rightarrow e^{-2 \alpha\phi} g_{\mu\nu}\,,\qquad
\varphi = e^{ \beta \phi}\,,\label{conformscaling}
\end{equation}
with
\begin{equation}
\alpha = -\fft{1}{\sqrt{2(D-1)(D-2)}}\,,\qquad \beta=
\sqrt{\fft{D-2}{2(D-1)}}\,.\label{alphabeta}
\end{equation}
The Lagrangian (\ref{gaugegenlag}) becomes
\begin{equation}
e^{-1}{\cal L}= \varphi \Big (R + (D-1)(D-3) g^2 \Big) + \varphi^3 \Big(-\ft14 {\cal F}_\2^2 + g^2(D-1)\Big)
-\ft14 \varphi^{-1} F_\2^2\,. \label{gaugegenfr}
\end{equation}
The solution (\ref{cbh1}) now becomes
\begin{eqnarray}
ds_D^2&=&-{\cal H}^{-1} H^{-1}\, h\, dt^2
+ H^{\fft{2}{D-3}}
\Big(\fft{dr^2}{h} + r^2 d\Omega_{D-2,\ep}^2\Big)\,,\cr 
{\cal F}_\2&=& \sqrt{\ep + \mu \tilde q^{-1}}\,dt\wedge
d{\cal H}^{-1}\,,\quad F_\2=\sqrt{\fft{D-1}{D-3}}\,\sqrt{\ep + \mu q^{-1}}\,dt\wedge dH^{-1}\,,\cr 
h &=& \ep -\fft{\mu}{r^{D-3}} + g^2 r^2 {\cal H}
H^{\fft{D-1}{D-3}}\,,\qquad \varphi = \sqrt{\fft{\cal
H}{H}}\,,\cr 
{\cal H} &=& 1 +
\fft{\tilde q}{r^{D-3}}\,,\qquad H=1 +
\fft{q}{r^{D-3}}\,.\label{cbh2}
\end{eqnarray}
The thermodynamics remain unchanged.

   Let us consider the case with $\tilde q=0$.  The solution becomes
\begin{eqnarray}
ds_D^2&=&-H^{-1}\, h\, dt^2
+ H^{\fft{2}{D-3}}
\Big(\fft{dr^2}{h} + r^2 d\Omega_{D-2,\ep}^2\Big)\,,\cr 
F_\2&=&\sqrt{\fft{D-1}{D-3}}\,\sqrt{\ep + \mu q^{-1}}\,dt\wedge dH^{-1}\,,\qquad
\varphi = H^{-\fft12}\,,\cr
h &=& \ep -\fft{\mu}{r^{D-3}} + g^2 r^2
H^{\fft{D-1}{D-3}}\,,\qquad H=1 +
\fft{q}{r^{D-3}}\,.\label{cbh3}
\end{eqnarray}
The mass of the solution is given by
\begin{equation}
M=\fft{\omega}{16\pi}\Big( (D-2)\mu + (D-3) \ep q\Big)\,.
\end{equation}
Now let us examine the global structure of the solution.  The geometry has two asymptotic regions at $r\rightarrow -\infty$ and $r\rightarrow +\infty$ and is completely free of curvature singularities for nonzero $q$. In particular, the scalar curvature invariants go as $1/(r^{D-3}+q)^n$ for some $n$ and are finite everywhere.
Whether or not there are horizons present depends on the relative values of the parameters $\mu$ and $q$.  If we have
\begin{equation}
\mu> \mu_0\equiv g^2 q^{\fft{D-1}{D-3}}\,,
\end{equation}
then $h$ must have a largest root at some $r=r_0$ and the solution describes a nonsingular black hole for which the horizons are located at $r=\pm r_0$. However, for positive $q$, $h$ can never have a double root.  Nevertheless, there exists an extremal limit of $\mu=\mu_0$ for which the near-horizon geometry is AdS$_2\times \Omega^{D-2,\ep}$. For $\mu<\mu_0$, the solution describes a wormhole \cite{susyfr}.

    It is instructive to express the solution using the new radial coordinate
$z=r^{(D-3)/2}$:
\begin{eqnarray}
ds_D^2 &=& -\fft{h}{z^2 + q} dt^2 + (z^2 +
q^2)^{\fft{2}{D-3}} \Big( \fft{4dz^2}{(D-3)^2 h} +
d\Omega_{D-2,\ep}^2\Big)\,,\cr 
A_\1 &=&\sqrt{\fft{D-1}{D-3}}\,\sqrt{\ep + \mu q^{-1}}\,\fft{q}{z^2 + q} dt\,,\qquad \varphi=\fft{|z|}{z^2 + q}\,,\cr
h&=& \ep z^2 -\mu + \mu_0 (z^2 q^{-1} +1 )^{\fft{D-1}{D-3}}\,.
\end{eqnarray}
Then one can see that the metric is smooth everywhere. The two asymptotic regions $z=\pm \infty$ are shielded by two horizons for $\mu>\mu_0$, which coalesce at $z=0$ when $\mu=\mu_0$. Thus, the solution describes a nonsingular black hole for $\mu>\mu_0$, an extremal black hole with the near-horizon geometry AdS$_2\times \Omega^{D-2,\ep}$ for $\mu=\mu_0$, and a wormhole for $\mu<\mu_0$.

\end{document}